\documentclass[12pt,letterpaper]{article}

\usepackage{jheppub,psfrag,amsbsy}
\newcommand{\bea}{\begin{eqnarray}}
\newcommand{\eea}{\end{eqnarray}}

\newcommand{\be}{\begin{equation}}
\newcommand{\ee}{\end{equation}}

\title{\Large Pohlmeyer reduction and Darboux transformations\\ in Euclidean worldsheet $AdS_3$}

\author{Georgios Papathanasiou}
\affiliation{Institute for Fundamental Theory\\
Department of Physics, University of Florida, Gainesville FL 32611}
\emailAdd{georgios@ufl.edu}


\arxivnumber{1203.3460 [hep-th]}

\abstract{Pohlmeyer reduction has been instrumental both in the program for computing gluon scattering amplitudes at strong coupling, and more recently in the progress towards semiclassical three-point correlators of heavy operators in $AdS/CFT$. After a detailed review of the method, we combine it with Darboux and Crum transformations in order to obtain a class of string solutions corresponding to an arbitrary number of kinks and breathers of the elliptic sinh-Gordon equation. We also use our construction in order to identify the previously found dressed giant gluon with the single breather solution.}

\begin{document}

\maketitle

\section{Introduction}

Anti-de Sitter space plays a central role in all instances of gauge/string duality (see \cite{Maldacena:2011ut} for a recent review). Closed strings propagating on this background are dual to gauge theory operators with nontrivial charges under the spacetime symmetry, and hence may share certain universal features with operators of nonsupersymmetric gauge theories. The GKP folded rotating string \cite{Gubser:2002tv} is precisely one of the first examples beyond the supergravity approximation, found to correspond to an operator with scaling dimension $\Delta$, high Lorentz spin $S$ and classical twist $\Delta-S=2$, exhibiting logarithmic scaling
\begin{equation}
\Delta-S=f(\lambda)\log S+\mathcal{O}(\log^0 S)\,,
\end{equation}
similarly to high spin operators present in QCD \cite{Gross:1974cs,Georgi:1951sr}\footnote{For an analysis of the fluctuation spectrum around the GKP string at any value of the 't Hooft coupling $\lambda$, providing the scaling dimensions of more high spin operators, see \cite{Basso:2010in}.}.

As the dual gauge theory lives on the boundary of $AdS$, open strings whose endpoints trace a closed boundary curve can also be put to correspondence with gauge theory observables, in particular Wilson loops \cite{Rey:1998ik,Maldacena:1998im}. The leading order expectation value of the Wilson loop on the curve is related to the minimal area swept by the classical string, and more recently it has been discovered that if the curve consists of straight lightlike segments, this is equivalent to planar gluon scattering amplitudes at strong coupling \cite{Alday:2007hr}\footnote{Each lightlike segment is then equal to the momentum of an external gluon.}. In the latter reference the four gluon amplitude was considered, and agreement was found with the all-loop prediction \cite{Anastasiou:2003kj,Bern:2005iz}, conjectured on the basis of the weak-coupling structure of the amplitude. Further continuation of this program has been very successful, first by showing the incompleteness of the aforementioned conjecture when the number of gluons is large \cite{Alday:2007he}, and then by computing the 8-point amplitude in special two-dimensional kinematics \cite{Alday:2009yn}, the 6-point amplitude in any kinematics \cite{Alday:2009dv}, and finally the general $n$-point amplitude in any kinematics \cite{Alday:2010vh}.

These results have been made possible with the help of a systematic procedure for eliminating all constraints of the string sigma model and ending up with equations of motion for the remaining physical degrees of freedom, known as Pohlmeyer reduction. Initially developed for the $O(n)$ model \cite{Pohlmeyer:1975nb}, where the reduced dynamics are described by the integrable sine-Gordon equation and its generalizations\footnote{For a good introduction on integrable two-dimensional field theories of the sine-Gordon type, and their pulselike wave solutions known as solitons, see \cite{Scott:1973eg}.}, it was later generalized to the non-compact $O(n,1)$ and $O(n,2)$ models \cite{Barbashov:1980kz,DeVega:1992xc, deVega:1992yz,Larsen:1996gn} of conformal gauge strings moving on de Sitter and anti-de Sitter spacetime. Particularly the case of $AdS_3$ is equivalent to the sinh-Gordon equation (an analytic continuation of the sine-Gordon equation), and has been studied very extensively the past few years in the series of papers \cite{Jevicki:2007aa,Jevicki:2008mm,Jevicki:2009uz,Jevicki:2010yt}. There, new spiky string solutions were constructed, a relation between spikes and sinh-Gordon kinks was established, and a map of the sigma model Lax pair to the sinh-Gordon Lax pair was found, after a spectral parameter was introduced.

Pohlmeyer reduction in $AdS_3$ has also been an integral part of the ongoing effort \cite{Janik:2011bd,Kazama:2011cp} for the computation of holographic three-point correlators of heavy operators, whose charges scale as $\mathcal{O}(\sqrt{\lambda})$. We refer the reader to the bibliography of the latter two references for developments in the study of 3-point correlation functions, which is currently a very active area of research. This owes to the great progress achieved in addressing the $AdS/CFT$ spectral problem due to integrability (see \cite{Beisert:2010jr} for a review), which completely determines the 2-point functions, and hence offers hope about exact solvability also in this context\footnote{For recent progress on the role of integrability in the computation of Wilson loops, see also \cite{Drukker:2012de,Correa:2012hh}.}. We should also note that for the holographic calculation of both Wilson loops and correlation functions to have the prescribed behavior at the $AdS$ boundary, the worldsheet has to be Euclidean (although in the case of correlation functions the boundary curve has to additionally collapse to a point). Other aspects of Pohlmeyer reduction in $AdS$ space have been considered in \cite{Miramontes:2008wt,Dorn:2009kq,Sakai:2009ut,Dorn:2009gq,Ryang:2009ay,Sakai:2010eh,Dorey:2010iy,Ishizeki:2011bf,Dorey:2011gr}.

In light of the potential application open strings may have in the study of the two aforementioned topics, in this note we complement Pohlmeyer reduction in Euclidean worldsheet $AdS_3$ with a method for constructing new solutions based on Darboux and Crum transformations \cite{Matveev1991,Andreev199558,springerlink:10.1023/A:1007583627973,1999physics...8019C}. Given a known solution of the reduced Lax pair equations, the Darboux transformation maps it algebraically to a new solution, very similarly to how the dressing method \cite{Zakharov:1973pp,Spradlin:2006wk,Kalousios:2006xy} works directly at the level of the sigma model. The Crum transformation is then equivalent to an $N$-fold Darboux transformation, and allows us to superpose any number of kinks and breathers on top of an initial `vacuum' solution.

Apart from introducing the general method in the context of $AdS_3$ strings for the first time, we also demonstrate its applicability by focusing on the specific example where the initial sinh-Gordon solution is equal to zero. In this manner, we obtain new multi-kink string configurations with noncompact Euclidean worldsheets\footnote{As we explain in section \ref{sec:reconstruction}, this involves the nontrivial step of solving an inverse scattering problem, where the sinh-Gordon solutions determine the potential. For this reason, only the single-kink string configuration with Euclidean worldsheet had previously appeared in the literature \cite{Sakai:2009ut}.}, analogous to the ones found in the Minkowskian case \cite{Jevicki:2009uz}. We are confident we will be able to apply it to other more complicated vacua including the GKP string as well, and work in this direction is currently underway.

This paper is organized as follows. Chapter \ref{sec_pohlmeyer_reduction} contains a pedagogical review of Pohlmeyer reduction on Euclidean worldsheet $AdS_{3}$, and sets up our notations. We start chapter \ref{sec:newsols} with a rederivation of the simplest `vacuum' solution in section \ref{sec:vacuum}, and use it in section \ref{sec:darboux} as input in order to algebraically construct the single kink (or antikink) sinh-Gordon field and corresponding string coordinate with the help of a Darboux transformation. In section \ref{sec_crum} we define the Crum transformation,  and apply it in the particular case at hand for the determination of the novel $N$-kink string solution. We further generalize to breathers in section \ref{sec_breathers}, and show that the single breather coincides with the dressed giant gluon solution found in \cite{Jevicki:2007pk}. We present our conclusions in chapter \ref{sec:conclusions}, and also include an appendix where we prove that our solutions are real for any $N$.

\section{Review of Pohlmeyer reduction on $AdS_{3}$}\label{sec_pohlmeyer_reduction}

In this chapter, we start by reviewing how the string equations of motion and Virasoro constraints imply the sinh-Gordon equations for the physical degrees of freedom, and how to inversely go from a known sinh-Gordon solution to a corresponding string configuration by means of a system of linear differential equations, the Lax pair. Finally we outline how to make this system amenable to the integrable methods we will subsequently use, by a redefinition and the introduction of an additional (spectral) parameter. As we mentioned in the introduction, most of these steps have been analyzed in \cite{Jevicki:2007aa,Jevicki:2008mm,Jevicki:2009uz,Jevicki:2010yt}, however our presentation will be mostly based on the notation of \cite{Alday:2009yn}, which is best suited for the Euclidean worldsheet.

\subsection{Derivation of the (generalized) sinh-Gordon equation}\label{sec_shG}

$AdS_{3}$ space can be parametrized in terms of $R^{2,2}$ embedding
coordinates $\vec{Y}\equiv(Y_{-1},Y_{0},Y_{1},Y_{2})$ with metric
$\eta^{\mu\nu}=\text{diag}(--++)$, obeying the constraint
\begin{equation}
\vec{Y}\cdot\vec{Y}\equiv\eta^{\mu\nu}Y_{\mu}Y_{\nu}=-(Y_{-1})^{2}-(Y_{0})^{2}+(Y_{1})^{2}+(Y_{2})^{2}=-1\,,\label{eq:quadratic_constraint}
\end{equation}
whose invariance under $SO(2,2)$ rotations reflects the corresponding
space isometries%
\footnote{We will be using the same conventions for any other $R^{2,2}$ vectors
and their scalar products. Notice in particular that we express a
vector in terms of its covariant (index-down) components. %
}. Then, strings in $AdS_{3}$ can be described by the sigma model
action,
\begin{equation}
S=\frac{T}{2}\int d\tau d\sigma[\sqrt{-\det(h_{cd})}h^{ab}\partial_{a}\vec{Y}\cdot\partial_{b}\vec{Y}+\Lambda(\vec{Y}\cdot\vec{Y}+1)]\label{eq:Action}
\end{equation}
 where $T$ is the string tension, $h^{ab}$ the worldsheet metric,
and the Lagrange multiplier $\Lambda$ is enforcing the aforemenioned
constraint. In what follows, we will pick the Euclidean conformal
gauge $h^{ab}=\text{diag}(1,1)$, and then switch to lightcone worldsheet
coordinates
\begin{align}
z & \equiv\frac{1}{2}(\sigma+i\tau) & \sigma & =\bar{z}+z & \partial\equiv\partial_{z} & =\partial_{\sigma}-i\partial_{\tau} & \partial_{\sigma} & =\frac{1}{2}(\bar{\partial}+\partial)\,,\\
\bar{z} & \equiv\frac{1}{2}(\sigma-i\tau) & \tau & =i(\bar{z}-z) & \bar{\partial}\equiv\partial_{\bar{z}} & =\partial_{\sigma}+i\partial_{\tau} & \partial_{\tau} & =\frac{1}{2i}(\bar{\partial}-\partial)\,.
\end{align}
which are clearly related by complex conjugation. In this case, the
equations of motion and Virasoro constraints that follow from (\ref{eq:Action})
become
\begin{equation}
\partial\bar{\partial}\vec{Y}-(\partial\vec{Y}\cdot\bar{\partial}\vec{Y})\vec{Y}=0\,,\quad\partial\vec{Y}\cdot\partial\vec{Y}=\bar{\partial}\vec{Y}\cdot\bar{\partial}\vec{Y}=0\,.\label{eq:sigma_model_EOM}
\end{equation}

Pohlmeyer reduction allows us to eliminate both Lagrange multiplier
and Virasoro constraints, so as to obtain equations of motion for
the remaining physical degrees of freedom. To this end, we define
the following $SO(2,2)$ scalar (invariant) quantities%
\footnote{In particular, in our definition $\alpha$ is half the one defined
in \cite{Alday:2009yn}.}
\begin{equation}
\begin{aligned}\alpha(z,\bar{z}) & \equiv\log(\frac{1}{2}\partial\vec{Y}\cdot\bar{\partial}\vec{Y})\,,\\
p(z,\bar{z}) & \equiv-\frac{1}{2}\vec{N}\cdot\partial^{2}\vec{Y}\,,\\
\bar{p}(z,\bar{z}) & \equiv\frac{1}{2}\vec{N}\cdot\bar{\partial}^{2}\vec{Y\,},
\end{aligned}
\label{eq:Pohlmeyer_scalars}
\end{equation}
 where the vector $\vec{N}$, defined in turn by
\begin{equation}
N_{\mu}\equiv\frac{e^{-\alpha}}{2}\epsilon_{\mu\nu\rho\sigma}Y^{\nu}\partial Y^{\rho}\bar{\partial}Y^{\sigma}\,,
\end{equation}
can be shown to obey
\begin{equation}
\bar{N}_{\mu}=-N_{\mu}\,,\quad\vec{N}\cdot\vec{Y}=\vec{N}\cdot\partial\vec{Y}=\vec{N}\cdot\bar{\partial}\vec{Y}=0\,,\quad\vec{N}\cdot\vec{N}=1\,.\label{eq:N_identities}
\end{equation}
The first equality above implies that $\vec{N}$ is purely imaginary,
and hence $p$ and $\bar{p}$ are related by complex conjugation.

In order to derive the equations of motion for (\ref{eq:Pohlmeyer_scalars}),
we will need to look at the derivatives of the vectors defining them,
and a convenient way to do so is by introducing the following moving
reference frame%
\footnote{We have chosen to take half the sum and difference of the vectors $q_2$ and $q_3$ in \cite{Alday:2009yn} for our basis so as to make it orthonormal, see also eq. (9) in \cite{Jevicki:2008mm} and eq. (B.1) in \cite{Neveu:1977cr}.
}
\begin{equation}
q_{1}=\vec{Y}\,,\quad q_{2}=\frac{e^{-\alpha/2}}{2}(\bar{\partial}\vec{Y}-\partial\vec{Y})\,,\quad q_{3}=\frac{e^{-\alpha/2}}{2}(\bar{\partial}\vec{Y}+\partial\vec{Y})\,,\quad q_{4}=\vec{N\,,}\label{eq:q_basis}
\end{equation}
whose scalar products follow from the previous definitions as
\begin{equation}
q_{i}\cdot q_{j}=\left(\begin{array}{cccc}
-1 & 0 & 0 & 0\\
0 & -1 & 0 & 0\\
0 & 0 & 1 & 0\\
0 & 0 & 0 & 1
\end{array}\right)_{ij}=\eta_{ij}\,.\label{eq:basis_products}
\end{equation}
The above relation shows that the linear transformations $q'_{i}\equiv R_{i}^{\phantom{{1}}m}q_{m}$
mapping to a new basis which preserves orthonormality, $q'_{i}\cdot q'_{j}=\eta_{ij}$,
will also be $SO(2,2)$ rotations, namely there is an internal $SO(2,2)$
symmetry transforming the $q_{i}$ as vectors with respect to the
$i$ index.

Taking the derivatives of the basis vectors in (\ref{eq:q_basis}),
and then using (\ref{eq:quadratic_constraint}), (\ref{eq:sigma_model_EOM})-(\ref{eq:N_identities}),
it is possible to reexpress them in terms of the basis vectors. In
particular $\partial\bar{\partial}\vec{Y}$ may be substituted from
(\ref{eq:sigma_model_EOM}), whereas for $\partial^{2}\vec{Y},\,\partial\vec{N}$
and their conjugates, it's easier to perform a substitution in terms
of the simpler but non-orthogonal basis
\begin{equation}
\vec{X}=a_{1}\vec{Y}+a_{2}\partial\vec{Y}+a_{3}\bar{\partial}\vec{Y}+a_{4}\vec{N}\,
\end{equation}
and then determine the coefficients from the products%
\footnote{In this manner we find $\partial^{2}\vec{Y}=\partial\alpha\partial\vec{Y}-2p\vec{N}$,
$\partial\vec{N}=pe^{-\alpha}\bar{\partial}Y$, and their conjugates.%
}
\begin{equation}
\begin{aligned}a_{1} & =-\vec{X}\cdot Y\,, & a_{3} & =\frac{1}{2}e^{-\alpha}\vec{X}\cdot\partial\vec{Y}\,,\\
a_{2} & =\frac{1}{2}e^{-\alpha}\vec{X}\cdot\bar{\partial}\vec{Y\,,} & a_{4} & =\vec{X}\cdot\vec{N}\,.
\end{aligned}
\end{equation}
The result we obtain is
\begin{equation}
\begin{aligned}\bar{\partial}q_{i} & =A_{ij}q_{j}\,, & A & =\left(\begin{array}{cccc}
0 & e^{\alpha/2} & e^{\alpha/2} & 0\\
-e^{\alpha/2} & 0 & \frac{1}{2}\bar{\partial}\alpha & e^{-\alpha/2}\bar{p}\\
e^{\alpha/2} & \frac{1}{2}\bar{\partial}\alpha & 0 & e^{-\alpha/2}\bar{p}\\
0 & e^{-\alpha/2}\bar{p} & -e^{-\alpha/2}\bar{p} & 0
\end{array}\right)\,,\\
\partial q_{i} & =B_{ij}q_{j}\,, & B & =\left(\begin{array}{cccc}
0 & -e^{\alpha/2} & e^{\alpha/2} & 0\\
e^{\alpha/2} & 0 & -\frac{1}{2}\partial\alpha & e^{-\alpha/2}p\\
e^{\alpha/2} & -\frac{1}{2}\partial\alpha & 0 & -e^{-\alpha/2}p\\
0 & e^{-\alpha/2}p & e^{-\alpha/2}p & 0
\end{array}\right)\,,
\end{aligned}
\label{eq:SO_laxpair}
\end{equation}
for which the compatibility condition $\partial(\bar{\partial}q)=\bar{\partial}(\partial q)$
implies $\partial A-\bar{\partial}B+[A,B]=0,$ and whose components
finally yield the equations of motion for the scalar quantities
\begin{align}
\partial\bar{\mbox{\ensuremath{\partial}}}\alpha-2e^{\alpha}+2p\bar{p}e^{-\alpha} & =0\,,\label{eq:shG_general}\\
\bar{\partial}p=\partial\bar{p} & =0\,.\label{eq:ppbar_EOM}
\end{align}
The first line is the generalized sinh-Gordon equation, and the second
line imposes that $p=p(z)$ is holomorphic and $\bar{p}=\bar{p}(\bar{z})$
antiholomorphic. The fact that $p$ ($\bar{p})$ can otherwise be
an arbitrary function of $z$ ($\bar{z}),$is a consequence of the
invariance of the equations of motion and Virasoro contraints (\ref{eq:sigma_model_EOM})
under conformal transformations on the worldsheet, $z\to z'=f(z)$
($\bar{z}\to\bar{z}'=f(\bar{z})$). As far as the transformation properties
of the quantities (\ref{eq:Pohlmeyer_scalars}) are concerned, if
we denote them with primes in the ($z',\bar{z'}$) frame, $\alpha'=\log(\partial'\vec{Y}\cdot\bar{\partial'}\vec{Y}/2)$
etc, it's easy to show that they will be related to the ones in the
($z,\bar{z}$) frame by
\begin{gather}
\alpha'=\alpha-\log(\frac{\partial z'}{\partial z}\frac{\partial\bar{z}'}{\partial\bar{z}})\,\\
p'=(\frac{\partial z}{\partial z'})^{2}p\,,\quad\bar{p}'=(\frac{\partial\bar{z}}{\partial\bar{z}'})^{2}\bar{p}\,.
\end{gather}
In particular this implies that starting with any given solution,
we can always perform a transformation
\begin{equation}
\begin{aligned}z' & =\int\sqrt{p(z)}dz\\
\bar{z}' & =\int\sqrt{\bar{p}(\bar{z})}d\bar{z}
\end{aligned}
\,\Rightarrow\begin{aligned}\alpha' & =\alpha-\frac{1}{2}\log(p\bar{p})\\
p' & =\bar{p}'=1
\end{aligned}
\,,
\end{equation}
so that $\alpha'$ obeys the usual sinh-Gordon equation
\begin{equation}
\partial'\bar{\mbox{\ensuremath{\partial}}'}\alpha'-4\sinh\alpha'=0\,.\label{eq:sinh_Gordon}
\end{equation}
Note however that the new variables will have square root branch cuts
where $p$ becomes zero, so that the information in $p$ in (\ref{eq:shG_general})-(\ref{eq:ppbar_EOM})
is encoded in the analytic structure of the $z'$ plane in (\ref{eq:sinh_Gordon}).

It is interesting to mention that the sinh-Gordon equation also appears in the study of scattering in Matrix String Theory \cite{Giddings:1998yd}. In the latter context, the relevant solutions of the equation are obtained by a limiting process where the number of kinks becomes both infinite and continuous \cite{Bonora:2002ay}.

\subsection{Reconstruction of the string coordinates}\label{sec:reconstruction}

The procedure we described in the previous section reduces the constrained
system of equations for the embedding $AdS_{3}$ coordinates (\ref{eq:sigma_model_EOM})
into the simpler set of equations (\ref{eq:shG_general})-(\ref{eq:ppbar_EOM})
of the generalized sinh-Gordon model. Due to the nonlocal nature of
(\ref{eq:Pohlmeyer_scalars}) however, knowledge of the sinh-Gordon
scalars does not immediately yield the string coordinates, and here
we will review how the latter can be reconstructed from the former.

In principle, one could substitute the sinh-Gordon scalars in the
$SO(2,2)$ Lax pair equations (\ref{eq:SO_laxpair}), and solve for
the $q_{i}.$ What is very particular about $SO(2,2)$ however, is
that it can be reduced to a direct product of two $SL(2)$ (or equivalently
$SO(2,1)$ or $SU(1,1)$ ) subgroups. We can thus exploit this feature
to similarly decompose (\ref{eq:SO_laxpair}) into two copies of much
simpler $SL(2)$ Lax pairs, and solve these instead.

We start by mapping all target space vectors into an equivalent bispinor
representation
\begin{equation}
X{}_{a\dot{a}}\equiv X_{\mu}\sigma_{\phantom{}a\dot{a}}^{\mu}=\left(\begin{array}{cc}
X_{-1}+X_{2} & X_{1}-X_{0}\\
X_{1}+X_{0} & X_{-1}-X_{2}
\end{array}\right)_{a\dot{a}}\,,\label{eq:Mu_to_a_dota}
\end{equation}
where $X_{\mu}\to(q_{i})_{\mu}$ and $\sigma_{\phantom{}a\dot{a}}^{\mu}\equiv(I_{2\times2},-i\sigma^{2},\sigma^{1},\sigma^{3})_{a\dot{a}}$
with $\sigma^{i}$ the usual Pauli matrices. In this formulation,
the invariance of the vector norm is translated as the invariance
of the matrix determinant, or more generally
\begin{equation}
\vec{X}\cdot\vec{Y}=-\frac{1}{2}X_{a\dot{a}}Y_{b\dot{b}}\epsilon^{ab}\epsilon^{\dot{a}\dot{b}},\label{eq:vector_product}
\end{equation}
with the standard convention $\epsilon^{12}=-\epsilon^{21}=-\epsilon_{12}=\epsilon_{21}=1$.

The same procedure can be carried out for the internal $SO(2,2)$
acting on the $q_{i}$,
\begin{equation}
W_{\alpha\dot{\alpha},a\dot{a}}\equiv\frac{1}{2}(q_{i})_{a\dot{a}}\tilde{\sigma}_{\phantom{}\alpha\dot{\alpha}}^{i}=\frac{1}{2}\left(\begin{array}{cc}
q_{1}-q_{4} & q_{3}+q_{2}\\
q_{3}-q_{2} & q_{1}+q_{4}
\end{array}\right)_{\alpha\dot{\alpha}}\,,\label{eq:W_definition}
\end{equation}
where $\tilde{\sigma}_{\phantom{}\alpha\dot{\alpha}}^{i}\equiv(I_{2\times2},i\sigma^{2},\sigma^{1},-\sigma^{3})_{\alpha\dot{\alpha}}$, and in the last equality we omitted the latin indices for brevity.
We may also translate the scalar products (\ref{eq:basis_products})
in this $SL(2)\times SL(2)$ bispinor notation if we multiply both
sides of the formula in question with $\tilde{\sigma}_{\phantom{}\alpha\dot{\alpha}}^{i}\tilde{\sigma}_{\phantom{}\beta\dot{\beta}}^{j}$,
and use (\ref{eq:vector_product}), (\ref{eq:W_definition}) and the
identity $\tilde{\sigma}_{\phantom{}\alpha\dot{\alpha}}^{i}\eta_{ij}\tilde{\sigma}_{\phantom{}\beta\dot{\beta}}^{j}=-\epsilon_{\alpha\beta}\epsilon_{\dot{\alpha}\dot{\beta}}$,
arriving at
\begin{equation}
\epsilon^{ab}\epsilon^{\dot{a}\dot{b}}W_{\alpha\dot{\alpha},a\dot{a}}W_{\beta\dot{\beta},b\dot{b}}=\epsilon_{\alpha\beta}\epsilon_{\dot{\alpha}\dot{\beta}}\,.\label{eq:q_i.q_j}
\end{equation}
We can similarly treat the completeness relation for the basis vectors
$q_{i}$%
\footnote{Namely the fact that we can write any $AdS_{3}$ vector as $X^{\mu}=\sum_{i}c^{i}(q_{i})^{\mu}$.%
}, $\eta^{ij}(q_{i})_{\mu}(q_{j})_{\nu}=\eta_{\mu\nu}$, in order to
obtain
\begin{equation}
\epsilon^{\alpha\beta}\epsilon^{\dot{\alpha}\dot{\beta}}W_{\alpha\dot{\alpha},a\dot{a}}W_{\beta\dot{\beta},b\dot{b}}=\epsilon_{ab}\epsilon_{\dot{a}\dot{b}}\,.\label{eq:completeness}
\end{equation}

We will now use the above formula to decompose $W$ into spinors that
transform irreducibly with respect to the $SL(2)$ subgroups of both
internal and spacetime $SO(2,2)$. Starting with $a=b$ and $\dot{a}=\dot{b}$,
(\ref{eq:completeness}) implies that the determinant of $W$ with
respect to the Greek indices is zero, and hence we can write it as
the tensor product of a left and a right spinor,
\begin{equation}
W_{\alpha\dot{\alpha},a\dot{a}}=\Psi_{\alpha,a\dot{a}}^{L}\Psi_{\dot{\alpha},a\dot{a}}^{R}\,,\label{eq:W_tensor_product}
\end{equation}
where by having the latin indices repeated without being summed, we
temporarily depart from a manifestly covariant formulation.

In what follows, we will think of the latin indices as labeling different spinors, and the Greek indices labeling the two different components of each spinor. When omitting the component index and looking at each spinor as a single object, we will use boldface, $\boldsymbol{\Psi}_{a\dot{a}}^{L}$, $\boldsymbol{\Psi}_{a\dot{a}}^{R}$ to make the distinction.

Given that the spinors have definite transformation properties with
respect to the internal $SL(2)_{L}\times SL(2)_{R}$ symmetry acting
on the Greek indices, we can define the following inner products in
each of the two subspaces,
\begin{equation}
\langle\boldsymbol{\chi}^{L},\boldsymbol{\psi}^{L}\rangle\equiv\epsilon^{\beta\alpha}\chi_{\alpha}^{L}\psi_{\beta}^{L}\,,\quad\langle\boldsymbol{\chi}^{R},\boldsymbol{\psi}^{R}\rangle\equiv\epsilon^{\dot{\beta}\dot{\alpha}}\chi_{\dot{\alpha}}^{R}\psi_{\dot{\beta}}^{R}\,,
\end{equation}
which are clearly antisymmetric in the exhange of the two spinors,
and furthermore it's easy to show that if
\begin{equation}
\langle\boldsymbol{\chi}^{L,R},\boldsymbol{\psi}^{L,R}\rangle=0\,\Rightarrow\chi_{\alpha}^{L,R}=\mu\psi_{\alpha}^{L,R}
\end{equation}
for some coefficient $\mu$, namely the two spinors are parallel.
In this notation, (\ref{eq:completeness}) may be rewritten as
\begin{equation}
\langle\boldsymbol{\Psi}_{a\dot{a}}^{L},\boldsymbol{\Psi}_{b\dot{b}}^{L}\rangle\langle\boldsymbol{\Psi}_{a\dot{a}}^{R},\boldsymbol{\Psi}_{b\dot{b}}^{R}\rangle=\epsilon_{ab}\epsilon_{\dot{a}\dot{b}}\,,\label{eq:completeness2}
\end{equation}
and by now looking at the $a\ne b$, $\dot{a}\ne\dot{b}$ cases we
deduce that
\begin{equation}
\langle\boldsymbol{\Psi}_{1\dot{1}}^{L,R},\boldsymbol{\Psi}_{2\dot{2}}^{L,R}\rangle,\langle\boldsymbol{\Psi}_{1\dot{2}}^{L,R},\boldsymbol{\Psi}_{2\dot{1}}^{L,R}\rangle\ne0\,.\label{eq:off_diagonal_products}
\end{equation}

Finally, considering the $a\ne b$, $\dot{a}=\dot{b}$ or $a=b$,
$\dot{a}\ne\dot{b}$ cases reveals that the only choice of inner products
we can set to zero without contradicting (\ref{eq:off_diagonal_products})
(up to trivially exchanging $L\leftrightarrow R$) is
\begin{equation}
\begin{aligned}\langle\boldsymbol{\Psi}_{a\dot{1}}^{L},\boldsymbol{\Psi}_{a\dot{2}}^{L}\rangle & =0\,\Rightarrow & \boldsymbol{\Psi}_{\alpha,a\dot{a}}^{L} & =c_{\dot{a}}^{L}\tilde{\boldsymbol{\Psi}}_{\alpha,a}^{L}\,,\\
\langle\boldsymbol{\Psi}_{1\dot{a}}^{R},\boldsymbol{\Psi}_{2\dot{a}}^{R}\rangle & =0\,\Rightarrow & \boldsymbol{\Psi}_{\dot{\alpha},a\dot{a}}^{R} & =c_{a}^{R}\tilde{\boldsymbol{\Psi}}_{\dot{\alpha},\dot{a}}^{R}\,,
\end{aligned}
\end{equation}
namely the dependence of the left (right) spinors on the dotted (undotted)
index can only appear as an overall multiplication factor, the remaining
part of the spinor being denoted with tilde. Then, in (\ref{eq:W_tensor_product})
the left factor can be reabsorbed by the right spinor and vice versa,
$\boldsymbol{\Psi}_{\alpha,a}^{L}\equiv c_{a}^{R}\tilde{\boldsymbol{\Psi}}_{\alpha,a}^{L}$,
$\boldsymbol{\Psi}_{\dot{\alpha},\dot{a}}^{R}\equiv c_{\dot{a}}^{L}\tilde{\boldsymbol{\Psi}}_{\dot{\alpha},\dot{a}}^{R}$,
so that $W$ can always be written as%
\footnote{The decomposition with respect ot the latin indices is also consistent
with $W_{\alpha\dot{\alpha},a\dot{a}}$ being lightlike spacetime
vectors, as can be seen from (\ref{eq:W_definition}) or (\ref{eq:q_i.q_j}).%
}
\begin{equation}
W_{\alpha\dot{\alpha},a\dot{a}}=\boldsymbol{\Psi}_{\alpha,a}^{L}\boldsymbol{\Psi}_{\dot{\alpha},\dot{a}}^{R}\,.\label{eq:W_decomposition}
\end{equation}
Now that we've decomposed $W$ in terms of spinors which transform
irreducibly under the spacetime symmetry as well, the separate index
structure in (\ref{eq:completeness}) or (\ref{eq:completeness2})
implies
\begin{equation}
\langle\boldsymbol{\Psi}_{a}^{L},\boldsymbol{\Psi}_{b}^{L}\rangle=c\epsilon_{ab}\,,\quad\langle\boldsymbol{\Psi}_{\dot{a}}^{R},\boldsymbol{\Psi}_{\dot{b}}^{R}\rangle=\frac{1}{c}\epsilon_{\dot{a}\dot{b}}\,,\label{eq:spinor_normalization}
\end{equation}
providing the freedom for two independent nontrivial rescalings of
the four spinors, which we will fix later.

We next turn to the decomposition of $SO(2,2)$ Lax pair equations
(\ref{eq:SO_laxpair}) into the corresponding equations for the spinors
$\boldsymbol{\Psi}^{L}$, $\boldsymbol{\Psi}^{R}$. Keeping the latin indices suppressed, and
treating $W$ as a matrix with respect to the Greek indices, we can
rewrite (\ref{eq:SO_laxpair}) the with the help of (\ref{eq:W_definition})
as
\begin{equation}
\begin{aligned}\partial W+B_{z}^{L}W+W(B_{z}^{R})^{T} & =0\,,\\
\bar{\partial}W+B_{\bar{z}}^{L}W+W(B_{\bar{z}}^{R})^{T} & =0\,,
\end{aligned}
\label{eq:W_EOM}
\end{equation}
where $(B^{L})_{\alpha}^{\phantom{}\beta}$, $(B^{R})_{\dot{\alpha}}^{\phantom{}\dot{\beta}}$
are the desired $SL(2)$ analogues of the Lax connections $A$, $B$
in (\ref{eq:SO_laxpair}),
\begin{equation}
\begin{aligned}B_{z}^{L} & =\left(\begin{array}{cc}
\frac{1}{4}\partial\alpha & -e^{\alpha/2}\\
-e^{-\alpha/2}p(z) & -\frac{1}{4}\partial\alpha
\end{array}\right) & B_{\bar{z}}^{L} & =\left(\begin{array}{cc}
-\frac{1}{4}\bar{\partial}\alpha & -e^{-\alpha/2}\bar{p}(\bar{z})\\
-e^{\alpha/2} & \frac{1}{4}\bar{\partial}\alpha
\end{array}\right)\,,\\
B_{z}^{R} & =\left(\begin{array}{cc}
-\frac{1}{4}\partial\alpha & e^{\alpha/2}p(z)\\
-e^{-\alpha/2} & \frac{1}{4}\partial\alpha
\end{array}\right) & B_{\bar{z}}^{R} & =\left(\begin{array}{cc}
\frac{1}{4}\bar{\partial}\alpha & -e^{-\alpha/2}\\
e^{\alpha/2}\bar{p}(\bar{z}) & -\frac{1}{4}\bar{\partial}\alpha
\end{array}\right)\,.
\end{aligned}
\label{eq:Lax_Pair_AM-1}
\end{equation}
In the matrix notation we employed for (\ref{eq:W_EOM}), and the boldface notation we introduced for the spinors, (\ref{eq:W_decomposition})
becomes $W=\boldsymbol{\Psi}^{L}(\boldsymbol{\Psi}^{R})^{T}$, and plugging this to the former
formulas we deduce
\begin{equation}
\begin{aligned}(\partial\boldsymbol{\Psi}^{L}+B_{z}^{L}\boldsymbol{\Psi}^{L})(\boldsymbol{\Psi}^{R})^{T}+\boldsymbol{\Psi}^{L}(\partial\boldsymbol{\Psi}^{R}+B_{z}^{R}\boldsymbol{\Psi}^{R})^{T} & =0\,,\\
(\bar{\partial}\boldsymbol{\Psi}^{L}+B_{\bar{z}}^{L}\boldsymbol{\Psi}^{L})(\boldsymbol{\Psi}^{R})^{T}+\boldsymbol{\Psi}^{L}(\bar{\partial}\boldsymbol{\Psi}^{R}+B_{\bar{z}}^{R}\boldsymbol{\Psi}^{R})^{T} & =0\,,
\end{aligned}
\end{equation}
whose most general solutions are%
\footnote{For example we can show this by taking the determinants of the above
matrix expressions, which imply that either the spinors or the left-
or right- hand side are linearly dependent, and then replace this
back in the expressions.%
}
\begin{equation}
\begin{aligned}\partial\boldsymbol{\Psi}^{L}+B_{z}^{L}\boldsymbol{\Psi}^{L} & =-\mu\boldsymbol{\Psi}^{L} & \partial\boldsymbol{\Psi}^{R}+B_{z}^{R}\boldsymbol{\Psi}^{R} & =\mu\boldsymbol{\Psi}^{R}\,,\\
\bar{\partial}\boldsymbol{\Psi}^{L}+B_{\bar{z}}^{L}\boldsymbol{\Psi}^{L} & =-\nu\boldsymbol{\Psi}^{L} & \bar{\partial}\boldsymbol{\Psi}^{R}+B_{\bar{z}}^{R}\boldsymbol{\Psi}^{R} & =\nu\boldsymbol{\Psi}^{R}\,,
\end{aligned}
\label{eq:SL2_Lax_equations}
\end{equation}
where the coefficients $\mu$, $\nu$ are for the moment general and
may depend on $z$, $\bar{z}$. Taking the compatibility conditions
$\bar{\partial}(\partial\boldsymbol{\Psi}^{L,R})=\partial(\bar{\partial}\boldsymbol{\Psi}^{L,R})$
leads to
\begin{equation}
(\bar{\partial}\mu+\partial\nu)I+(\partial B_{\bar{z}}^{L}-\bar{\partial}B_{z}^{L}+[B_{z}^{L},B_{\bar{z}}^{L}])=0\,,\quad-(\bar{\partial}\mu+\partial\nu)I+(\partial B_{\bar{z}}^{R}-\bar{\partial}B_{z}^{R}+[B_{z}^{R},B_{\bar{z}}^{R}])=0\,,
\end{equation}
and looking at the components we notice that we separately have to
satisfy $\bar{\partial}\mu+\partial\nu=0$ and%
\footnote{In component form we can also observe that the two flatness conditions
are not independent, since one is minus the transpose of the other
one.%
}
\begin{equation}
\partial B_{\bar{z}}^{L}-\bar{\partial}B_{z}^{L}+[B_{z}^{L},B_{\bar{z}}^{L}]=0\,,\quad\partial B_{\bar{z}}^{R}-\bar{\partial}B_{z}^{R}+[B_{z}^{R},B_{\bar{z}}^{R}]=0\,,\label{eq:B_flatness}
\end{equation}
which as expected also yield the generalized sinh-Gordon equations
(\ref{eq:shG_general})-(\ref{eq:ppbar_EOM}).

The connections $B^{L,R}$ and spinors $\boldsymbol{\Psi}^{L,R}$ are not uniquely
defined, as $SL(2)$ gauge transformations

\begin{equation}
\boldsymbol{\Psi}'=V\boldsymbol{\Psi},\quad B'_{z}=VB_{z}V^{-1}-\partial VV^{-1}\quad B'_{\bar{z}}=VB_{\bar{z}}V^{-1}-\bar{\partial}VV^{-1}\label{eq:SL2_gauge_transformations}
\end{equation}
holding for both $L,\, R$ variables, and in general with different
$V\to V^{L,R}$, leave the equations (\ref{eq:SL2_Lax_equations})
invariant. We can thus use them to eliminate the inhomogeneous term,
in particular by picking $V^{L}=fI$, $V^{R}=-fI$, such that $\mu=\partial(\log f)$,
$\nu=\bar{\partial}(\log f)$. In other words, we can without loss
of generality replace (\ref{eq:SL2_Lax_equations}) with the homogeneous
auxiliary linear problems

\begin{equation}
\begin{aligned}\partial\boldsymbol{\Psi}^{L}+B_{z}^{L}\boldsymbol{\Psi}^{L} & =0 & \partial\boldsymbol{\Psi}^{R}+B_{z}^{R}\boldsymbol{\Psi}^{R} & =0\,,\\
\bar{\partial}\boldsymbol{\Psi}^{L}+B_{\bar{z}}^{L}\boldsymbol{\Psi}^{L} & =0 & \bar{\partial}\boldsymbol{\Psi}^{R}+B_{\bar{z}}^{R}\boldsymbol{\Psi}^{R} & =0\,,
\end{aligned}
\label{eq:SL2_Lax_FINAL}
\end{equation}
where $B^{L,R}$ are again given by (\ref{eq:Lax_Pair_AM-1}).

Each of these problems will have two linearly independent solutions,
which are precisely the $\Psi_{\alpha,a}^{L}$ for $a=1,2$ and $\Psi_{\dot{\alpha},\dot{a}}^{R}$
for $\dot{a}=1,2$. Indeed any special linear transformation of the
two solutions will also be a solution and will leave the inner products
(\ref{eq:spinor_normalization}) invariant, and thus we can identify
the indices labeling the solutions with the spacetime $SL(2)$ indices.
This also shows that different choices of independent solutions or
integration constants amount to different choices of a spacetime frame,
and we have the freedom to select them in the most convenient fashion.

Finally, by acting on (\ref{eq:spinor_normalization}) with either
$\partial$ or $\bar{\partial}$ and using (\ref{eq:SL2_Lax_FINAL})
we can show that the inner products are independent of $z$, $\bar{z}$
and hence $c$ is a constant. It then follows straightforwardly that
the nontrivial rescalings of the spinors which change the value of
$c$ correspond to Lorentz boosts, and since these can be accounted
for by transforming the latin indices of the spinors, we can make
the most symmetric choice $c=1$.

So the coordinates of a string corresponding to a solution of the
sinh-Gordon equation (\ref{eq:shG_general}-\ref{eq:ppbar_EOM}) may
be obtained by replacing the solution in the Lax matrices (\ref{eq:Lax_Pair_AM-1}),
solving the Lax pair equations (\ref{eq:SL2_Lax_FINAL}) for the spinors
under the aforementioned normalization condition, and replacing them
in
\begin{equation}
Y_{a\dot{a}}=\left(\begin{array}{cc}
Y_{-1}+Y_{2} & Y_{1}-Y_{0}\\
Y_{1}+Y_{0} & Y_{-1}-Y_{2}
\end{array}\right)_{a\dot{a}}=(q_{1})_{a\dot{a}}=W_{1\dot{1},a\dot{a}}+W_{2\dot{2},a\dot{a}}=(\boldsymbol{\Psi}_{a}^{L})^{T}(\boldsymbol{\Psi}_{\dot{a}}^{R})\,,\label{eq:Y_from_LR}
\end{equation}
where we used (\ref{eq:q_basis}), (\ref{eq:Mu_to_a_dota}), (\ref{eq:W_definition}),
(\ref{eq:W_decomposition}) and in the last equality we again employed
matrix notation with respect to the Greek indices.

\subsection{Spectral parameter and a convenient gauge}\label{sec:spectram_param}

The final crucial ingredient which facilitates the solution of integrable
systems, is the introduction of an additional variable in the auxiliary
linear problem (\ref{eq:SL2_Lax_FINAL}), the spectral parameter $\lambda$,
such that its compatibility condition still gives rise to the equations of motion (\ref{eq:shG_general})-(\ref{eq:ppbar_EOM}) for any
value of $\lambda$. In particular, we can see that if we generalize
the left connection (\ref{eq:Lax_Pair_AM-1}) as
\begin{equation}
\hat{B}_{z}(\lambda)=\left(\begin{array}{cc}
\frac{1}{4}\partial\alpha & -\frac{1}{\lambda}e^{\alpha/2}\\
-\frac{1}{\lambda}e^{-\alpha/2}p(z) & -\frac{1}{4}\partial\alpha
\end{array}\right)\,,\qquad\hat{B}_{\bar{z}}(\lambda)=\left(\begin{array}{cc}
-\frac{1}{4}\bar{\partial}\alpha & -\lambda e^{-\alpha/2}\bar{p}(\bar{z})\\
-\lambda e^{\alpha/2} & \frac{1}{4}\bar{\partial}\alpha
\end{array}\right)\,,\label{eq:Lax_Pair_AM}
\end{equation}
it also obeys (\ref{eq:B_flatness}). In fact, both connections (\ref{eq:Lax_Pair_AM-1})
may be obtained from the above formula by specializing to specific
values of $\lambda$, plus a gauge transformation of the type (\ref{eq:SL2_gauge_transformations})
for the right problem,

\begin{equation}
B_{z}^{L}=\hat{B}_{z}(1)\,,\quad B_{z}^{R}=U\hat{B}_{z}(i)U^{-1}\,,\quad U=\frac{1}{\sqrt{2}}\left(\begin{array}{cc}
0 & 1+i\\
-1+i & 0
\end{array}\right),\label{eq:U}
\end{equation}
with identical relations for $z\to\bar{z}$. Consecutively if $\hat{\boldsymbol{\Psi}}(\lambda)$
is the solution of the linear problem with the connections (\ref{eq:Lax_Pair_AM}),
then the left and right spinors may be obtained as
\begin{equation}
\boldsymbol{\Psi}^{L}=\hat{\boldsymbol{\Psi}}(1),\quad\boldsymbol{\Psi}^{R}=U\hat{\boldsymbol{\Psi}}(i)\,.\label{eq:U_spinor_transformation}
\end{equation}

For our purposes, it will be convenient to perform an additional gauge
transformation

\begin{equation}
\Psi(\lambda)=V\hat{\Psi}(\lambda),\quad B_{z}=V\hat{B}_{z}V^{-1}-\partial VV^{-1}\quad\hat{B}_{\bar{z}}=VB_{\bar{z}}V^{-1}-\bar{\partial}VV^{-1}\label{eq:V_transformation}
\end{equation}
with $V=\text{diag}(e^{\mbox{\ensuremath{\alpha}/4}},e^{\mbox{-\ensuremath{\alpha}/4}})$,
such that
\begin{equation}
B_{z}(\lambda)=\left(\begin{array}{cc}
0 & -\frac{1}{\lambda}e^{\alpha}\\
-\frac{1}{\lambda}e^{-\alpha}p(z) & 0
\end{array}\right)\,,\qquad B_{\bar{z}}(\lambda)=\left(\begin{array}{cc}
-\frac{1}{2}\bar{\partial}\alpha & -\lambda\bar{p}(\bar{z})\\
-\lambda & \frac{1}{2}\bar{\partial}\alpha
\end{array}\right)\,.\label{eq:Lax_pair_new}
\end{equation}

Gathering everything together, for a given known solution of the sinh-Gordon
equation (\ref{eq:shG_general})-(\ref{eq:ppbar_EOM}), we replace
its functional form in the Lax pairs (\ref{eq:Lax_pair_new}) and
solve the auxiliary problem
\begin{equation}
\partial\boldsymbol{\Psi}(\lambda)+B_{z}(\lambda)\boldsymbol{\Psi}(\lambda)=0\,,\qquad\bar{\partial}\boldsymbol{\Psi}(\lambda)+B_{\bar{z}}(\lambda)\boldsymbol{\Psi}(\lambda)=0,\label{eq:Linear_problem}
\end{equation}
where the general solution $\boldsymbol{\Psi}(\lambda)$ is spanned
by a superposition of two linearly independent $SL(2)$ spinors $\boldsymbol{\Psi}_{a}(\lambda)$
\begin{equation}
\boldsymbol{\Psi}(\lambda)=\sum_{a=1}^{2}\boldsymbol{\Psi}_{a}(\lambda),\label{eq:Psi_superposition}
\end{equation}
whose components are denoted by $\Psi_{\alpha,a}$ for $\alpha=1,2$. Each of the $\boldsymbol{\Psi}_{a}(\lambda)$ includes an integration
constant, and we normalize them such that
\begin{equation}
\langle\boldsymbol{\Psi}_{a},\boldsymbol{\Psi}_{b}\rangle=\epsilon^{\beta\alpha}\Psi_{\alpha,a}\Psi_{\beta,b}=\epsilon_{ab}\,.\label{eq:Psi_normalization}
\end{equation}
Finally, the corresponding $AdS_{3}$ string solution will be obtained
from the above spinors as%
\footnote{This formula follows from (\ref{eq:Y_from_LR}), (\ref{eq:U_spinor_transformation})
and (\ref{eq:V_transformation}), after we note that $(V^{-1})^{T}UV^{-1}=U$.%
}
\begin{equation}
Y_{a\dot{a}}=\left(\begin{array}{cc}
Y_{-1}+Y_{2} & Y_{1}-Y_{0}\\
Y_{1}+Y_{0} & Y_{-1}-Y_{2}
\end{array}\right)_{a\dot{a}}=\boldsymbol{\Psi}_{a}^{T}(1)U\boldsymbol{\Psi}_{\dot{a}}(i)\,,\label{eq:Y_groupelement}
\end{equation}
where we treat $\dot{a}$ just as another index labeling the same
two linearly independent solutions in (\ref{eq:Psi_superposition}),
and $U$ is given in (\ref{eq:U}).

\section{New solutions via Darboux and Crum transformations}\label{sec:newsols}
In this chapter, we will combine the general auxiliary linear problem and coordinate reconstruction formula (\ref{eq:Lax_pair_new})-(\ref{eq:Y_groupelement}) with the method of Darboux and Crum transformations \cite{Matveev1991,Andreev199558,springerlink:10.1023/A:1007583627973,1999physics...8019C}, in order to produce new string solutions corresponding to multikink configurations of the elliptic (Euclidean) sinh-Gordon, or sinh-Laplace, equation.

In particular we will focus on the case where $p, \bar p$ are smooth and nonzero on the $z$ plane\footnote{This clearly excludes polynomial functions, but for example includes exponentials of polynomial functions.}, and choose a convenient worldsheet frame that sets them to $p=\bar p=1$ without changing the analytic structure on the plane.

\subsection{Warmup: The giant gluon vacuum}\label{sec:vacuum}

The simplest solution of (\ref{eq:shG_general}) with $p=\bar p=1$ is clearly the `vacuum' $\alpha=0$. For this choice, the associated spinor of the auxiliary
problem (\ref{eq:Lax_pair_new})-(\ref{eq:Linear_problem}), can be found to be
\begin{equation}
\boldsymbol{\Psi}(\lambda)=\sum_{a=1}^{2}\boldsymbol{\Psi}_{a}(\lambda)=c_{1}e^{-(z/\lambda+\bar{z}\lambda)}\left(\begin{array}{c}
-1\\
1
\end{array}\right)+c_{2}e^{(z/\lambda+\bar{z}\lambda)}\left(\begin{array}{c}
1\\
1
\end{array}\right)\;,\label{eq:vacuum_spinor}
\end{equation}
and a particular choice compatible with (\ref{eq:Psi_normalization})
which makes (\ref{eq:Y_groupelement}) real is $c_{1}=-\bar{c}_{2}=(1+i)/2$.
With this choice, we can describe the two linearly independent solutions
$\boldsymbol{\Psi}_{a}$ jointly as
\begin{equation}
\boldsymbol{\Psi}_{a}(\lambda)=\frac{-(-1)^{a}+i}{2}\, e^{(-1)^{a}(z/\lambda+\bar{z}\lambda)}\left(\begin{array}{c}
(-1)^{a}\\
1
\end{array}\right)\,,\label{eq:indep_Phi}
\end{equation}
and plugging this in (\ref{eq:Y_groupelement}) yields
\begin{equation}
Y_{a\dot{a}}=\frac{1}{\sqrt{2}}\left(\begin{array}{cc}
e^{-\sigma-\tau} & e^{-\sigma+\tau}\\
-e^{\sigma-\tau} & e^{\sigma+\tau}
\end{array}\right)\,\Rightarrow\vec{Y}=\frac{1}{\sqrt{2}}\left(\begin{array}{c}
\cosh(\sigma+\tau)\\
-\cosh(\sigma-\tau)\\
-\sinh(\sigma-\tau)\\
-\sinh(\sigma+\tau)
\end{array}\right)\,,\label{eq:giantgluon}
\end{equation}
which can be easily shown to satisfy
\begin{equation}
Y_{0}^{2}-Y_{-1}^{2}=Y_{1}^{2}-Y_{2}^{2}\,.
\end{equation}

Namely it is the cusp solution first found in \cite{Kruczenski:2002fb},
and then used by \cite{Alday:2007hr} in the context of gluon scattering
amplitudes at strong coupling. In the latter paper a four gluon solution
was also obtained, termed as `giant gluon' in \cite{Jevicki:2007pk},
which is related to (\ref{eq:giantgluon}) by simple rotations of
the embedding coordinates ($SO(2,2)$ isometries of $AdS_{3}$) ,
\begin{equation}\label{eq:AM_fourcusp}
\vec{Y}'=\left(\begin{array}{cccc}
\frac{1}{\sqrt{2}} & -\frac{1}{\sqrt{2}} & 0 & 0\\
\frac{1}{\sqrt{2}} & \frac{1}{\sqrt{2}} & 0 & 0\\
0 & 0 & -\frac{1}{\sqrt{2}} & -\frac{1}{\sqrt{2}}\\
0 & 0 & \frac{1}{\sqrt{2}} & -\frac{1}{\sqrt{2}}
\end{array}\right)\vec{Y}=\left(\begin{array}{c}
\cosh\sigma\cosh\text{\ensuremath{\tau}}\\
\sinh\text{\ensuremath{\sigma}}\sinh\tau\\
\sinh\sigma\cosh\tau\\
\cosh\text{\ensuremath{\sigma}}\sinh\tau
\end{array}\right)\,.
\end{equation}

\subsection{Darboux transformations and the single-kink string}\label{sec:darboux}

Of course the string configuration of the previous section has been well known in the literature, however in this section we will use it as input for a Darboux transformation, which will generate a different solution corresponding to a single sinh-Gordon (anti)kink.

The form of the new Lax pair (\ref{eq:Lax_pair_new}) we introduced is precisely necessary for the application of this method, as a second differentiation of $\bar{z}$ equation yields separate equations for each component of the spinor $\boldsymbol{\Psi}(\lambda)\equiv(\psi,\phi)$,
\begin{equation}\label{eq:sG_schroedinger}
\begin{aligned}
\bar{\partial}^{2}\psi-\lambda^{2}\psi & =(\frac{1}{2}\bar{\partial}^{2}\alpha+\frac{1}{4}(\bar{\partial}\alpha)^{2})\psi,\\
\bar{\partial}^{2}\phi-\lambda^{2}\phi & =(-\frac{1}{2}\bar{\partial}^{2}\alpha+\frac{1}{4}(\bar{\partial}\alpha)^{2})\phi,
\end{aligned}
\end{equation}
in which case we can take advantage of the following theorem (see \cite{WADATIMiki:1975-02-25,Matveev1991} for more information).

\noindent {\bf Darboux's theorem.} Consider the time-independent Schroedinger equation,
\begin{equation}
-\partial_{x}^{2}\psi+u\psi=-\lambda^{2}\psi,\label{eq:schroedinger}
\end{equation}
 for a known potential $u(x)$, where $-\lambda^{2}$ plays the role
of energy. If $\psi^{1}(x)$ a particular solution of (\ref{eq:schroedinger})
for $\lambda=\lambda_{1}$, and $\psi(x,\lambda)$ an arbitrary solution, then the function obtained by the (Darboux) transformation
\begin{equation}
\psi'=\frac{W(\psi^{1},\psi)}{\psi^{1}}=\frac{\psi^{1}\partial_x\psi-\partial_x\psi^{1}\psi}{\psi^{1}}\label{eq:Darboux_spinor}
\end{equation}
 where $W(\psi^{1},\psi)$ the Wronskian determinant, satisfies a
Schroedinger equation with the same energy eigenvalue%
\footnote{Hence the transformation is `isospectral'.%
}, but for a different potential,
\begin{equation}
\begin{gathered}
-\partial_{x}^{2}\psi'+u'\psi'=-\lambda^{2}\psi',\\
u'=u-2\partial_{x}^{2}\log\psi^{1}.
\end{gathered}
\end{equation}
 Clearly (\ref{eq:sG_schroedinger}) are of the form (\ref{eq:schroedinger})
with the corresponding potentials given on the right hand side. So
for a known sinh-Gordon solution $\alpha$ and its associated spinor for a particular value of the spectral parameter $\boldsymbol{\Psi}^1=(\psi^1,\phi^1)$,
the Darboux transformation generates a new solution $\alpha'$ via
the relation of the potentials,
\begin{equation}
\begin{aligned}
\frac{1}{2}\bar{\partial}^{2}\alpha'+\frac{1}{4}(\bar{\partial}\alpha')^{2} & =\frac{1}{2}\bar{\partial}^{2}\alpha+\frac{1}{4}(\bar{\partial}\alpha)^{2}-2\bar{\partial}^{2}\log\psi^{1},\\
-\frac{1}{2}\bar{\partial}^{2}\alpha'+\frac{1}{4}(\bar{\partial}\alpha')^{2} & =-\frac{1}{2}\bar{\partial}^{2}\alpha+\frac{1}{4}(\bar{\partial}\alpha)^{2}-2\bar{\partial}^{2}\log\phi^{1}.
\end{aligned}
\end{equation}
 Subtracting the two formulas and integrating%
\footnote{The linear and constant term are set to zero by the boundary condition
$\alpha\to0$ as $|\bar{z}|\to\infty$.%
}, we obtain a direct transformation for the $\alpha's$,
\begin{equation}
\alpha'-\alpha=2\log\phi^{1}-2\log\psi^{1}\,,\Rightarrow e^{\alpha'}=e^{\alpha}\left(\frac{\phi^{1}}{\psi^{1}}\right)^{2}.\label{eq:Darboux_alpha}
\end{equation}
 Furthermore, the Darboux transformation also gives us the associated
spinors (\ref{eq:Darboux_spinor}), which we can write as
\begin{equation}
\boldsymbol{\Psi}'(\lambda)=\left(\begin{array}{c}
\psi'\\
\phi'
\end{array}\right)=\lambda\left(\begin{array}{c}
\phi\\
\psi
\end{array}\right)-\lambda_{1}\left(\begin{array}{c}
\frac{\phi^{1}}{\psi^{1}}\psi\\
\frac{\psi1}{\phi^{1}}\phi
\end{array}\right),\label{eq:Darboux_Phi}
\end{equation}
up to a normalization factor which we'll fix shortly.

Let us now analyze what kind of new sinh-Gordon solutions we obtain by dressing
the vacuum $\alpha=0$, if we take $\boldsymbol{\Psi}^1=\boldsymbol{\Psi}(\lambda_1)$ from (\ref{eq:vacuum_spinor}), but with the integration constants unspecified,
\begin{equation}
e^{\alpha'}=\left(\frac{c_{2}e^{u_{1}}+c_{1}e^{-u_{1}}}{c_{2}e^{u_{1}}-c_{1}e^{-u_{1}}}\right)^{2}=\left(\frac{1+\frac{c_{2}}{c_{1}}e^{2u_{1}}}{1-\frac{c_{2}}{c_{1}}e^{2u_{1}}}\right)^{2}\,,\,u_{1}=(z/\lambda_{1}+\bar{z}\lambda_{1})\,.
\end{equation}
 Evidently, choosing any of the integration constants to zero, does
not produce a new solution. Furthermore, demanding that $\alpha'$ is real restricts $\lambda_1$ to be a pure phase, $\bar \lambda_1=1/\lambda_1$, and $c_{1}/c_{2}$ to be real. We can thus absorb its absolute value in the exponential, which simply shifts the profile parameter by a constant $k_{1}=\frac{1}{2}\log|c_{2}/c_{1}|$, and for positive or negative sign of the ratio, we obtain the sinh-Gordon
antikink $\alpha'_{+}$ or kink $\alpha'_{-}$ solution respectively
\begin{equation}
\alpha'_{\text{\text{+}}}=\log\coth^{2}u_{1}^{\prime},\quad\alpha'_{-}=\log\tanh^{2}u_{1}^{\prime}\,,\quad u_{1}^{\prime}=z/\lambda_{1}+\bar{z}\lambda_{1}+k_{1}\,.
\end{equation}
 We can also rewrite the vacuum spinors that lead to them as
\begin{equation}
\boldsymbol{\Psi}_{+}^{1}=\left(\begin{array}{c}
\sinh u_{1}^{\prime}\\
\cosh u_{1}^{\prime}
\end{array}\right)\:,\quad\boldsymbol{\Psi}_{-}^{1}=\left(\begin{array}{c}
\cosh u_{1}^{\prime}\\
\sinh u_{1}^{\prime}
\end{array}\right)\,,\label{eq:shG_kink_spinors}
\end{equation}
 up to an overall constant, which cancels out in (\ref{eq:Darboux_Phi}).
Replacing $\psi_{1},\,\phi_{1}$ from the above formula, and $\phi,\,\psi$
for each of the linearly independent solutions (\ref{eq:indep_Phi}),
we thus obtain the corresponding dressed spinors
\begin{equation}
\boldsymbol{\Psi}'_{a}(\lambda)=\frac{-(-1)^{a}+i}{2((-1)^{a}\lambda-\lambda_{1})}e^{(-1)^{a}(z/\lambda+\bar{z}\lambda)}\left(\begin{array}{c}
\lambda-(-1)^{a}\lambda_{1}(\coth u'_{1})^{\pm1}\\
(-1)^{a}\lambda-\lambda_{1}(\tanh u'_{1})^{\pm1}
\end{array}\right)\,,
\end{equation}
where plus or minus in the exponents of the hyperbolic functions corresponds
to the antikink or kink respectively, and the overall normalization
factor appearing in front, has been chosen such that for $u'_{1}\to\infty$,
$\boldsymbol{\Psi}'_{a}(\lambda)$ reduces to the linearly independent solutions
of the vacuum spinors (\ref{eq:indep_Phi}), and hence
also respects (\ref{eq:Psi_normalization}).

Then, with the help of
(\ref{eq:Y_groupelement}), we can write the coordinates of the single
spike solution as
\begin{equation}
Y_{a\dot{a}}=\frac{1}{\sqrt{2}}\left(\begin{array}{cc}
e^{-\sigma-\tau}\frac{\lambda_{1}^{2}+(1+i)(\coth u'_{1})^{\pm1}+i}{(\lambda_{1}+1)(\lambda_{1}+i)} & e^{-\sigma+\tau}\frac{\lambda_{1}^{2}+(1-i)(\coth u'_{1})^{\pm1}-i}{(\lambda_{1}+1)(\lambda_{1}-i)}\\
-e^{\sigma-\tau}\frac{\lambda_{1}^{2}-(1-i)(\coth u'_{1})^{\pm1}-i}{(\lambda_{1}-1)(\lambda_{1}+i)} & e^{\sigma+\tau}\frac{\lambda_{1}^{2}-(1+i)(\coth u'_{1})^{\pm1}+i}{(\lambda_{1}-1)(\lambda_{1}-i)}
\end{array}\right)\,,
\end{equation}
for which it is an easy task to check that it is indeed real, and
that it similarly reduces to the vacuum solution (\ref{eq:giantgluon})
for $u'_{1}\to\infty$. We can further simplify the above relation and make its reality manifest, by replacing $\lambda_1=e^{i p_1}$, in which case it becomes
\begin{equation}\label{eq:single_kink_string}
Y_{a\dot{a}}=\frac{1}{2}\left(\begin{array}{cc}
e^{-\sigma-\tau}\frac{(\coth u'_{1})^{\pm1} + \sqrt{2} \sin(p_1 + \frac{\pi}{4})}{(1 + \sqrt{2} \sin(p_1 + \frac{\pi}{4})} & e^{-\sigma+\tau}\frac{(\coth u'_{1})^{\pm1} - \sqrt{2} \sin(p_1 - \frac{\pi}{4})}{(1 - \sqrt{2} \sin(p_1 - \frac{\pi}{4})}\\
-e^{\sigma-\tau}\frac{(\coth u'_{1})^{\pm1} + \sqrt{2} \sin(p_1 - \frac{\pi}{4})}{(1 + \sqrt{2} \sin(p_1 - \frac{\pi}{4})} & e^{\sigma+\tau}\frac{(\coth u'_{1})^{\pm1} - \sqrt{2} \sin(p_1 + \frac{\pi}{4})}{(1 - \sqrt{2} \sin(p_1 + \frac{\pi}{4})}
\end{array}\right)\,,
\end{equation}
with $u_1'=\sigma\cos p_1 +\tau\sin p_1+k_1 $.

For $p_1=\pi/4$, (\ref{eq:single_kink_string}) can be shown to coincide with the Sommerfield-Thorn solution \cite{Sommerfield:2008hu} for positive hyperbolic cotangent power (antikink), and the Berkovits-Maldacena solution \cite{Berkovits:2008ic} for negative hyperbolic cotangent power\footnote{In particular for the latter case, a worldsheet conformal transformation $\sigma+\tau\to \tau$ and $\sigma-\tau\to\sigma$, which also takes $p=1\to p'=-i/2$, maps our solution to the conformal gauge form presented in \cite{Jevicki:2010yt}.} (kink), whereas the general $p_1$ solution was found by a limiting procedure in \cite{Sakai:2009ut}. It is important to emphasize that once the `vacuum' solution of the previous section (\ref{eq:indep_Phi}) has been determined, our method yields both kink and antikink solutions in a completely algebraic fashion.

\subsection{Crum transformations and the multi-kink string}\label{sec_crum}

Although the method of Darboux transformations we presented in the previous section may be used recursively in order to produce more string configurations, expressions tend to get lengthier and without any particular pattern at each iteration. Instead, here we will employ a generalization due to Crum \cite{1999physics...8019C} (see also \cite{Matveev1991} for a modern treatment), which solves the recursion of the $N$-fold transformation and directly produces multi-kink solutions in a simple form, with all the symmetry structure manifest. The reader who is interested in the final result may jump to equations (\ref{eq:N-Yaa})-(\ref{eq:N-phi/psi}).

\vspace{6pt}

\noindent {\bf Crum theorem.} For the Schroedinger equation (\ref{eq:schroedinger}) with given
potential $u(x)$, let $\psi^{1},\ldots,\psi^{N}$ be solutions for
specific values of the spectral parameter $\lambda=\lambda_{1},\ldots,\lambda_{N}$,
and $\psi=\psi(\lambda)$ solution for arbitrary spectral parameter.
If we define the Wronskian determinant of $k$ functions as
\begin{equation}
W(f_{1},\ldots,f_{k})\equiv\det(\frac{d^{i-1}f_{j}}{dx^{i-1}})\,,i,j=1,\ldots,k\,,
\end{equation}
then the function
\begin{equation}
\psi(\lambda_{1},\ldots,\lambda_{N};\lambda)\equiv\frac{W(\psi^{1},\ldots,\psi^{N},\psi)}{W(\psi^{1},\ldots,\psi^{N})}\,,
\end{equation}
satisfies (\ref{eq:schroedinger}) but for $u(x)$ replaced by a new
potential
\begin{equation}
u(\lambda_{1}\ldots,\lambda_{N};x)=u(x)-2\partial_{x}^{2}\log W(\psi^{1},\ldots,\psi^{N})\,.
\end{equation}
Applying this to (\ref{eq:sG_schroedinger}), similarly to the Darboux
transformation, we obtain that the $N$-th dressed solution of the
elliptic sinh-Gordon equation $\alpha(\lambda_{1},\ldots\lambda_{N})$,
and its associated spinor $\boldsymbol{\Psi}(\lambda_{1},\ldots,\lambda_{N};\lambda)$
are given by
\begin{equation}
\begin{aligned}e^{\alpha(\lambda_{1},\ldots\lambda_{N})} & =e^{\alpha}\left(\frac{W(\phi^{1},\ldots,\phi^{N})}{W(\psi^{1},\ldots,\psi^{N})}\right)^{2}\,,\\
\boldsymbol{\Psi}(\lambda_{1},\ldots,\lambda_{N};\lambda) & \equiv\left(\begin{array}{c}
\psi(\lambda_{1},\ldots,\lambda_{N};\lambda)\\
\phi(\lambda_{1},\ldots,\lambda_{N};\lambda)
\end{array}\right)=\left(\begin{array}{c}
\frac{W(\psi^{1},\ldots,\psi^{N},\psi)}{W(\psi^{1},\ldots,\psi^{N})}\\
\frac{W(\phi^{1},\ldots,\phi^{N},\phi)}{W(\phi^{1},\ldots,\phi^{N})}
\end{array}\right)\,.
\end{aligned}
\label{eq:Crum_transformation}
\end{equation}
The above theorem may be regarded as the reason for the appearance of determinant formulas representing multikink solutions in all integrable systems solvable by some variant of the inverse scattering method.

Returning to our problem at hand, we may use the Lax pair equation (\ref{eq:Linear_problem}) for $\bar \partial$ in order to eliminate all derivatives from the spinor components in (\ref{eq:Crum_transformation}), and in fact it turns out that
inside the Wronskian, we can simply replace\footnote{We remind the reader that we have made the choice $p=\bar{p}=1$.} \cite{Andreev199558}
\begin{equation}\label{eq:derivative_substitution}
\begin{aligned}\bar{\partial}^{2k+1}\psi^{i} & \to\lambda_{i}^{2k+1}\phi^{i} & \bar{\partial}^{2k}\psi^{i} & \to\lambda_{i}^{2k}\psi^{i}\,,\\
\bar{\partial}^{2k+1}\phi^{i} & \to\lambda_{i}^{2k+1}\psi^{i} & \bar{\partial}^{2k}\phi^{i} & \to\lambda_{i}^{2k}\phi^{i}\,,
\end{aligned}
\end{equation}
and similarly for $\phi,\,\psi$. This is because (\ref{eq:Linear_problem}) allows us to write the right-hand sides of
(\ref{eq:derivative_substitution}) as a sequence of $\ldots(\bar{\partial}-\bar{\partial}\alpha/2)(\bar{\partial}+\bar{\partial}\alpha/2)\ldots$
operators acting on $\psi^{i}$ or $\phi^{i}$, so that the highest
derivative term equals the the right-hand side plus lower derivative
terms, whose coefficients are the same for all values of $i$. Consequently
at each row of the determinant, the additional terms can be removed
by subtracting appropriate multiples of the previous rows. Hence we
can write
\begin{equation}\label{eq:Wronskian}
W(\psi^{1},\ldots,\psi^{N})=\left|\begin{array}{cccc}
\psi^{1} & \psi^{2} & \cdots & \psi^{N}\\
\lambda_{1}\phi^{1} & \lambda_{2}\phi^{2} & \cdots & \lambda_{N}\phi^{N}\\
\lambda_{1}^{2}\psi^{1} & \lambda_{2}^{2}\psi^{2} & \cdots & \lambda_{N}^{2}\psi^{N}\\
\vdots & \vdots & \cdots & \vdots
\end{array}\right|=\prod_{i=1}^{N}\psi^{i}\left|\begin{array}{cccc}
1 & 1 & \cdots & 1\\
\lambda_{1}\frac{\phi^{1}}{\psi^{1}} & \lambda_{2}\frac{\phi^{2}}{\psi^{2}} & \cdots & \lambda_{N}\frac{\phi^{N}}{\psi^{N}}\\
\lambda_{1}^{2} & \lambda_{2}^{2} & \cdots & \lambda_{N}^{2}\\
\vdots & \vdots & \cdots & \mathbf{\vdots}
\end{array}\right|\,,
\end{equation}
and similarly for $W(\phi^{1},\ldots,\phi^{N})$.

Having simplified the form of the Wronskians, there remains one additional
subtlety for their evaluation: Going to Euclidean worldsheet time
turns the sinh-Gordon equation into the sinh-Laplace equation%
\footnote{Namely the worldsheet D'Alembertian $-\frac{\partial^{2}}{\partial\tau^{2}}+\frac{\partial^{2}}{\partial\sigma^{2}}$
turns into a Laplacian $\frac{\partial^{2}}{\partial\tau^{2}}+\frac{\partial^{2}}{\partial\sigma^{2}}$.%
}, and as was shown in \cite{springerlink:10.1023/A:1007583627973},
in this case Darboux and Crum transformations map real solutions to
purely imaginary solutions and vice versa. Clearly a purely imaginary
sinh-Gordon solution is the same as a purely real sine-Gordon solution\footnote{The elliptic sine-Gordon equation was first studied in \cite{1977PhRvB..15.3353L}.},
so that for $\alpha_{i}\,,\beta_{i}\in R$, the transformations produce
the following two sequences that mix the two types of solutions,
\begin{equation}
\begin{aligned}\alpha_{0}\to & i\beta_{1}\to  \phantom{i}\alpha_{2}\to  i\beta_{3}\to  \ldots\,,&  (\frac{\partial^{2}}{\partial\tau^{2}}+\frac{\partial^{2}}{\partial\sigma^{2}})\alpha_{j}&=4\sinh\alpha_{j}\,,\\
i\beta_{0}\to & \phantom{i}\alpha_{1}\to  i\beta_{2}\to  \phantom{i}\alpha_{3}\to  \ldots\,,& (\frac{\partial^{2}}{\partial\tau^{2}}+\frac{\partial^{2}}{\partial\sigma^{2}})\beta_{j}&=4\sin\beta_{j}\,.
\end{aligned}
\label{eq:Darboux_sequences}
\end{equation}
Since strings in $AdS_{3}$ are reconstructed from sinh-Gordon solutions,
it is thus evident that for the $N$-th dressed solution we need to
use spinors associated to the single sinh-Gordon (anti)kink (\ref{eq:shG_kink_spinors})
if $N$ is odd, and spinors associated to the single sine-Gordon (anti)kink
if $N$ is even. Namely for $N$ even we should replace (\ref{eq:shG_kink_spinors}) with
\begin{equation}
\boldsymbol{\Psi}^i=\left(\begin{array}{c}
\psi^{i}\\
\phi^{i}
\end{array}\right)=\left(\begin{array}{c}
ie^{u_{i}^{\prime}}-e^{-u_{i}^{\prime}}\\
ie^{u_{i}^{\prime}}+e^{-u_{i}^{\prime}}
\end{array}\right)\,
\end{equation}
where $u_{i}^{\prime}=z/\lambda_{i}+\bar{z}\lambda_{i}+k_{i}$ and in this case we can describe both the kink (for $\text{Re}(\lambda_i)>0$) and antikink (for $\text{Re}(\lambda_i)<0$) simultaneously. We elaborate more on this choice, and on the validity of the sequences (\ref{eq:Darboux_sequences}), at the end of the appendix.

Finally, we need to normalize the spinors (\ref{eq:Crum_transformation})
in order to ensure that they too respect the normalization (\ref{eq:Psi_normalization}).
Similarly to the singly dressed case, we can achieve this by demanding
that the spinors (\ref{eq:Crum_transformation}) reduce to the vacuum
spinors (\ref{eq:indep_Phi}) when $\phi^{i}/\psi^{i}\to1$. In this
limit the Wronskians become identical to Vandermonde determinants,
\begin{equation}
\det(\lambda_{j}^{i-1})=\prod_{1\le i<j\le N}(\lambda_{j}-\lambda_{i})\,,i,j=1,\ldots N\,,
\end{equation}
and we can easily infer that the unnormalized spinors (\ref{eq:Crum_transformation})
are just $\prod_{i=1}^{N}((-1)^{a}\lambda-\lambda_{i})$ times the
vacuum spinors (\ref{eq:indep_Phi}).

Gathering all our results together, the embedding coordinates for
the $AdS_{3}$ string corresponding to the $N$-th dressed sinh-Gordon
solution will be given by
\begin{equation}
Y_{a\dot{a}}=\frac{1}{\sqrt{2}}\left[(1+i)\psi_{a}(\lambda_{1},\ldots\lambda_{N};1)\phi_{\dot{a}}(\lambda_{1},\ldots\lambda_{N};i)-(1-i)\phi_{a}(\lambda_{1},\ldots\lambda_{N};1)\psi_{\dot{a}}(\lambda_{1},\ldots\lambda_{N};i)\right]\,,\label{eq:N-Yaa}
\end{equation}
where the associated spinor components are given by%
\footnote{The formula is also valid for $a\to\dot{a}$.%
}
\begin{equation}
\begin{aligned}\psi_{a}(\lambda_{1},\ldots\lambda_{N};\lambda) & =\frac{-1+(-1)^{a}i}{2\prod_{i=1}^{N}((-1)^{a}\lambda-\lambda_{i})}e^{(-1)^{a}(z/\lambda+\bar{z}\lambda)}\frac{\det(\lambda_{j}^{i-1}r_{ij}^{+})_{N+1}}{\det(\lambda_{j}^{i-1}r_{ij}^{+})_{N}}\,,\\
\phi_{a}(\lambda_{1},\ldots\lambda_{N};\lambda) & =\frac{-1+(-1)^{a}i}{2\prod_{i=1}^{N}((-1)^{a}\lambda-\lambda_{i})}e^{(-1)^{a}(z/\lambda+\bar{z}\lambda)}\frac{\det(\lambda_{j}^{i-1}r_{ij}^{-})_{N+1}}{\det(\lambda_{j}^{i-1}r_{ij}^{-})_{N}}\,,
\end{aligned}
\label{eq:N-spinors}
\end{equation}
 with the determinants in the above formula corresponding to $(N+1)\times(N+1)$
matrices in the numerator, $N\times N$ matrices in the numerator%
\footnote{Namely $i,j$ run from 1 to $N+1$ in the numerator, and 1 to $N$
in the denominator.%
}, $\lambda_{N+1}\equiv\lambda$,
\begin{equation}
r_{ij}^{\pm}\equiv\frac{1}{2}(1\pm(-1)^{i})\frac{\phi^{j}}{\psi^{j}}+\frac{1}{2}(1\mp(-1)^{i})\,,
\end{equation}
 is equal to either $\phi^{j}/\psi^{j}$ or 1 depending on the sign
and value of $i$, and finally
\begin{equation}
\begin{aligned}\frac{\phi^{N+1}}{\psi^{N+1}} & =(-1)^{a}\,, & \frac{\phi^{j}}{\psi^{j}} & =\begin{cases}
\tanh(z/\lambda_{j}+\bar{z}\lambda_{j}+k_{j})\,, & \text{for a kink if\,}N\,\text{odd}\,,\\
\coth(z/\lambda_{j}+\bar{z}\lambda_{j}+k_{j}) & \text{for an antikink if}\, N\,\text{odd}\,,\\
\frac{ie^{2(z/\lambda_{j}+\bar{z}\lambda_{j}+k_{j})}+1}{ie^{2(z/\lambda_{j}+\bar{z}\lambda_{j}+k_{j})}-1} & \text{if }\, N\,\text{even}\,.
\end{cases}\, j=1\ldots N\,.\end{aligned}
\label{eq:N-phi/psi}
\end{equation}
In the above formula, we restrict $\text{Re}(\lambda_j)>0$ for $N$ odd\footnote{Considering $\text{Re}(\lambda_j<0)$ is equivalent to taking $\lambda_j\to -\lambda_j$, which we see may only change the Wronskians (\ref{eq:Wronskian}) up to an overall sign. This sign however cancels by squaring or taking their ratios in (\ref{eq:Crum_transformation}).}, whereas for $N$ even $\text{Re}(\lambda_j)>0$ corresponds to a kink and $\text{Re}(\lambda_j)<0$ to an antikink. As we have mentioned earlier in the text, (\ref{eq:N-Yaa})-(\ref{eq:N-phi/psi}) represents a solution corresponding to kinks and antikinks (without any bound states) when $\lambda_j=e^{i p_j}$ are phases.

The fact that the coordinates are ultimately expressed in terms of ratios of determinants with dimensions $N+1$ over $N$ is very reminiscent of the multisoliton solutions for sigma models with compact target spaces found in \cite{Kalousios:2010ne}. In the appendix we explicitly prove that they are real for any $N$.

\subsection{Breathers and the dressed giant gluon}\label{sec_breathers}

Here we extend the parameter space of the new $N$-kink solution, so as to also include oscillating lumps, or `breathers'. We consecutively use this extension in order to show the $AdS_3$ string confuguration first found in \cite{Jevicki:2007pk}, which we will henceforth call the `dressed giant gluon', is precisely a breather solution.

Let us start by focusing on the $N=2$ case, where if we let $r_j=\phi^j/\psi^j$ for compactness, the sinh-Gordon field (\ref{eq:Crum_transformation}) may be written with the help of (\ref{eq:Wronskian}),(\ref{eq:N-phi/psi}) as
\begin{equation}
e^{\alpha(\lambda_{1},\lambda_{2})} =\Big(\frac{\lambda_2 r_1-\lambda_1 r_2}{\lambda_2 r_2-\lambda_1 r_1}\Big)^2=\Big(\frac{\cosh(u_1+u_2) (\lambda_1-\lambda_2)-i \sinh(u_1-u_2) (\lambda_1+\lambda_2)}{\cosh(u_1+u_2) (\lambda_1-\lambda_2)+i \sinh(u_1-u_2) (\lambda_1+\lambda_2)}\Big)^2\,,
\label{eq:2kink}
\end{equation}
and its associated spinor can be massaged to
\begin{equation}
\boldsymbol{\Psi}_a(\lambda_{1},\lambda_{2};\lambda) =\prod_{i=1}^{2}((-1)^{a}\lambda-\lambda_{i})^{-1}\left(\begin{array}{cc}
\lambda^2+\lambda_1\lambda_2 \frac{\lambda_2 r_1-\lambda_1 r_2}{\lambda_2 r_2-\lambda_1 r_1}& \frac{1}{\lambda_2 r_2-\lambda_1 r_1}\\
\frac{1}{\lambda_2/r_2-\lambda_1/r_1}&\lambda^2+\lambda_1\lambda_2 \frac{\lambda_2 r_2-\lambda_1 r_1}{\lambda_2 r_1-\lambda_1 r_2}
\end{array}\right)\Psi_a(\lambda)\,,
\end{equation}
with $\Psi_a(\lambda)$ given by (\ref{eq:indep_Phi}).

If we now for a moment allow the $\lambda_i$ to be general complex numbers, and demand that the right-hand side of (\ref{eq:2kink}) is real and positive, it is straightforward to show that apart from the case $\bar\lambda_j=1/\lambda_j$, $\bar k_j=k_j$, which we had encountered so far, there exists precisely one more alternative restriction, $\bar\lambda_2=1/\lambda_1$, $\bar k_2=k_1$. This also leads to the profile parameters of the solitons being conjugate to each other $\bar u_2=u_1$, which is a feature common with the breathers of the ordinary (Minkowskian) sine and sinh- Gordon equation, although here the soliton parameters have a different range. A plot of the solution shows that it is a localized excitation consisting of two lumps oscillating together as if connected by a spring. This justifies the term `breather' used to describe them and allows their interpretation as a bound kink-antikink pair, see for example \cite{Rajaraman}.

We can further generalize these considerations, such that the $N$-th dressed sinh-Gordon field (\ref{eq:Crum_transformation}) and corresponding string configuration (\ref{eq:N-Yaa})-(\ref{eq:N-phi/psi}) describe for any $0\le n\le [N/2]$
\begin{enumerate}
\item $n$ breathers with $\bar\lambda_{j+1}=1/\lambda_j$, $\bar k_{j+1}=k_j$, $j=1,\ldots,n$,
\item $N-2n$ kinks or antikinks with $\bar\lambda_j=1/\lambda_j$, $\bar k_j=k_j$, $j=2n+1,\ldots,N$ .
\end{enumerate}
The proof of reality for these solutions goes along the lines of the $N$-kink case presented in the appendix, the only difference being that the determinant columns for each conjugate pair of soliton parameters will be conjugate to each other altogether. Hence complex conjugation of the determinant will give the same result as in the $N$-kink case up to a possible minus sign, which will cancel in the ratios of determinants in (\ref{eq:Crum_transformation}).

Having explored the full range of lump-like solutions over the sinh-Gordon vacuum $\alpha=0$, we next move on to interpret the $AdS_3$ dressed giant gluon found in \cite{Jevicki:2007pk} with the help of the dressing method \cite{Zakharov:1973pp} (see also \cite{Spradlin:2006wk,Kalousios:2006xy} for the first application in the $AdS/CFT$ context).

The motivation behind \cite{Jevicki:2007pk} stemmed from the then recently initiated program for computing gluon scattering amplitudes at strong coupling, by establishing the equivalence of the problem to computing the Euclidean worldsheet area of an open string whose ends form a light-like polygon on the boundary of $AdS$ space. In particular, for an $n$-point amplitude the boundary consists of $n$ light-like segments meeting at cusps, and after \cite{Alday:2007hr} studied the case $n=4$ in detail, there was need to find concrete realizations of higher point functions.

However a characteristic of the dressed giant gluon, that made its interpretation as an amplitude or more generally Wilson loop more intricate, was that it would reach the $AdS$ boundary an infinite amount of times and at finite values of the worldsheet coordinates. We quote the solution here, for a particular choice of its parameters,
\begin{equation}
Z_1\equiv Y_{-1}+i Y_0=\frac{1}{|\zeta|}\frac{\vec Y\cdot \vec N_1}{D}\,,\quad Z_2\equiv Y_{1}+i Y_2=\frac{1}{|\zeta|}\frac{\vec Y\cdot \vec N_2}{D},
\end{equation}
where $\vec Y$ represents the 4-cusp solution (\ref{eq:AM_fourcusp}), the vectors $\vec N_i$ are given by
\begin{equation}
\begin{aligned}
\vec{N}_1 &=
\begin{pmatrix}
- (\bar\zeta m \bar m - \zeta) \cosh(Z + \bar{Z})
+ i (\bar\zeta m \bar m + \zeta) \sinh(Z - \bar{Z}) \\
- (\zeta m \bar m + \bar\zeta) \sinh(Z - \bar{Z}) - i (\zeta m \bar m
- \bar\zeta) \cosh(Z + \bar{Z}) \cr
(\zeta - \bar\zeta) \bar{m} (\sinh(Z + \bar{Z}) - i
\cosh(Z - \bar{Z})) \\
(\zeta - \bar\zeta) m ( \cosh(Z - \bar{Z}) - i
\sinh(Z + \bar{Z}))
\end{pmatrix},\\
\vec{N}_2 &=  \begin{pmatrix}
- (\zeta - \bar\zeta) \bar{m} ( \sinh(Z + \bar{Z}) - i
\cosh(Z - \bar{Z}) ) \\
- (\zeta - \bar\zeta) m ( \cosh(Z - \bar{Z}) - i \sinh(Z + \bar{Z}))\cr
+ (\bar\zeta m \bar m - \zeta) \cosh(Z + \bar{Z}) - i (\bar\zeta
m \bar m + \zeta) \sinh(Z - \bar{Z})\\
+ (\zeta m \bar m + \bar\zeta) \sinh(Z - \bar{Z}) +
i (\zeta m \bar m - \bar\zeta) \cosh(Z + \bar{Z})
\end{pmatrix},
\end{aligned}
\end{equation}
the denominator responsible for the complicated boundary behavior is
\begin{equation}
D = (m \bar m - 1) \cosh(Z+\bar{Z}) - i (m \bar m + 1) \sinh(Z - \bar{Z})\,,
\end{equation}
and the soliton-like parameters entering in the above formulas are
\begin{equation}
Z=\frac{z}{m}+\bar z m\,,\quad\zeta=i \frac{1-m^2}{1+m^2}\,,
\end{equation}
with barred quantities denoting complex conjugatation.

Computing the Pohlmeyer-reduced sinh-Gordon field of this solution according to (\ref{eq:Pohlmeyer_scalars}) yields\footnote{We thank Chrysostomos Kalousios for pointing this out to us.}
\begin{equation}
e^\alpha=\frac{1}{2}\partial\vec Y\cdot \bar \partial\vec Y=\Big(\frac{\cosh(Z+\bar Z) (m \bar m-1)-i \sinh(Z-\bar Z) (m \bar m+1)}{\cosh(Z+\bar Z) (m \bar m-1)+i \sinh(Z-\bar Z) (m \bar m+1)}\Big)^2\,,
\end{equation}
which precisely agrees with (\ref{eq:2kink}) upon the identification $m=-\lambda_1=-1/\bar\lambda_2$, such that $Z=-u_1=-\bar u_2$ with $k_1=\bar k_2=0$, thus implying that the dressed giant gluon corresponds to a breather\footnote{This is similar to the solution obtained by dressing in Minkowskian worldsheet \cite{Jevicki:2007aa}.}.

We can now interpret the complicated behavior at the $AdS$ boundary, which corresponds to $|Z_i|\to \infty$, as follows. First, the soliton-like solutions of sinh-Gordon represent transitions between the two extrema of the field potential $\cosh\alpha$ at $\alpha\to \pm\infty$, so that they clearly reach infinite values (at the positions of the kinks) and have infinite energy. Second, the particular breather solution reaches $e^\alpha\to \infty$ at every oscillation, and also $|Z_i|\to \infty$ an infinite number of times due to their common denominator factor.

\section{Conclusions}\label{sec:conclusions}

In this paper we introduced a new method for constructing string solutions on $AdS_3$, based on the combination of Pohlmeyer reduction and Darboux and Crum transformations. As we reviewed in detail, Pohlmeyer reduction trades the
$AdS_3$ string equations of motion and Virasoro constraints with two sets of differential equations that have to be solved, first the sinh-Gordon equation (\ref{eq:shG_general})-(\ref{eq:ppbar_EOM}), and then its Lax pair equation (\ref{eq:Linear_problem}), where the sinh-Gordon solution is used as input. The advantage of our method is that for a single solution of the Lax pair, it yields an infinite class of new string configurations in a purely algebraic manner.

We focused on the case of Euclidean worldsheet, given its relevance in the computation on Wilson loops/gluon scattering amplitudes and 3-point correlators at strong coupling, but our framework can be applied equally well also for the Minkowskian worldsheet. We considered the simplest application where the initial sinh-Gordon field is zero, corresponding to the cusp solution \cite{Kruczenski:2002fb,Alday:2007hr}, and obtained spiky string solutions corresponding to kinks and breathers, similar to the ones in \cite{Jevicki:2009uz}. In particular our solutions are open strings with noncompact worldsheets, which reach the $AdS_3$ boundary at infinite and generically also at certain finite values of the worldsheet coordinates. Notice however that the single kink (-) solution in (\ref{eq:single_kink_string}) only reaches the boundary for $\tau,\sigma\to\pm \infty$, and thus it would be worthwhile to investigate whether there exists a region in the parameter space where this is valid for multikink solutions as well.

More importantly, it would be very interesting to employ our method to explore other classes of string solutions, including ones with compact worldsheets, and in particular we believe it is possible to obtain generalizations of the GKP string \cite{Gubser:2002tv}. Finally, it would be desirable for one to have a mechanism for generating string solutions directly at the level of the string sigma model, which can also be generalized to spaces of higher dimensionality, similar to the dressing method \cite{Zakharov:1973pp,Spradlin:2006wk,Kalousios:2006xy} for compact target spaces. As we discussed in the previous chapter, when applied to the the noncompact $AdS$ space, the dressing method superposes a bound kink-antikink pair at a time, rather than a single kink. Since Darboux transformations can be viewed as dressing the $SL(2)$ spinors whose bilinears make up the $AdS$ coordinates, they may provide information on how to correctly modify the dressing of the coordinates, so as to generate all solutions of a given class.

\section*{Acknowledgements}
We would like to thank Chrysostomos Kalousios for collaboration at the initial stages of this work. We are also grateful to Benjamin Basso and Pawel Caputa for enlightening discussions, and especially Marcus Spradlin, Charles Thorn and the anonymous referee of JHEP for several helpful comments on the manuscript. This work was supported in part by the Department of Energy under Grant No. DE-FG02-97ER-41029.

\appendix

\section{Reality of the $N$-th dressed solution}\label{appx_reality}

In this appendix, we will prove that the solution (\ref{eq:N-Yaa})-(\ref{eq:N-phi/psi}) we constructed is real for any $N$, and is mapped to a real sinh-Gordon field as well. This will also serve as a confirmation of the two distinct sequences of solutions produced by $N$ consecutive Darboux (or equivalently $N$-fold Crum) transformations (\ref{eq:Darboux_sequences}).

Let us first focus on the case where $N$ is odd. Because of (\ref{eq:N-phi/psi}), $\bar{r}_{ij}^{\pm}=r_{ij}^{\pm}$ is real, and since $\bar{\lambda}_{i}=1/\lambda_{i}$, we have
\begin{equation}
\begin{aligned}\overline{\det(\lambda_{j}^{i-1}r_{ij}^{\pm})}_{N} & =\det(\frac{1}{\lambda_{j}^{i-1}}r_{ij}^{\pm})_{N}=\prod_{j=1}^{N}\frac{1}{\lambda_{j}^{N-1}}\det(\lambda_{j}^{N-i}r_{ij}^{\pm})_{N}\\
 & =(-1)^{\frac{N-1}{2}}\prod_{j=1}^{N}\frac{1}{\lambda_{j}^{N-1}}\det(\lambda_{j}^{i-1}r_{ij}^{\pm})_{N}\,,
\end{aligned}
\label{eq:determinant_conjugation}
\end{equation}
where in the last equality of the first line we took out a factor $1/\lambda_{j}^{N-1}$ from
each row of the determinant. Moreover in the second line we inverted the
order of rows, picking up a possible minus sign depending on $N$,
and ending up with a multiple of the determinant we started with because $r_{N+1-i,j}^{\pm}=r_{ij}^{\pm}$. Similarly, we obtain
\begin{equation}
\begin{aligned}\overline{\det(\lambda_{j}^{i-1}r_{ij}^{\pm})}_{N+1} & =(-1)^{\frac{N+1}{2}}\prod_{j=1}^{N+1}\frac{1}{\lambda_{j}^{N}}\det(\lambda_{j}^{i-1}r_{ij}^{\mp})_{N+1}\,,\end{aligned}
\label{eq:determinat_conjugation2}
\end{equation}
where the main difference is that now the matrix is even-dimensional
and $r_{N+1-i,j}^{\pm}=r_{ij}^{\mp}$, or in other words 1 and $\phi^{j}/\psi^{j}$
exchange places. Taking the complex conjugate of $\psi_{a}(\lambda_{1},\ldots\lambda_{N};\lambda)$
in (\ref{eq:N-spinors}) and using the above relations, we thus find
\begin{equation}
\begin{aligned}\overline{\psi_{a}(\lambda_{1},\ldots\lambda_{N};\lambda)} & =\frac{-1-(-1)^{a}i}{2\prod_{i=1}^{N}(\frac{(-1)^{a}}{\lambda}-\frac{1}{\lambda_{i}})}e^{(-1)^{a}(z/\lambda+\bar{z}\lambda)}(-\prod_{j=1}^{N}\frac{1}{\lambda_{j}\lambda}\frac{\det(\lambda_{j}^{i-1}r_{ij}^{-})_{N+1}}{\det(\lambda_{j}^{i-1}r_{ij}^{+})_{N}})\\
 & =i\frac{-1+(-1)^{a}i}{2\prod_{i=1}^{N}((-1)^{a}\lambda-\lambda_{i})}e^{(-1)^{a}(z/\lambda+\bar{z}\lambda)}\frac{\det(\lambda_{j}^{i-1}r_{ij}^{-})_{N+1}}{\det(\lambda_{j}^{i-1}r_{ij}^{+})_{N}}\\
 & =i\frac{\det(\lambda_{j}^{i-1}r_{ij}^{-})_{N}}{\det(\lambda_{j}^{i-1}r_{ij}^{+})_{N}}\phi_{a}(\lambda_{1},\ldots\lambda_{N};\lambda)\,,
\end{aligned}
\label{eq:N-psi_conjugation}
\end{equation}
and by the same token
\begin{equation}
\overline{\phi_{a}(\lambda_{1},\ldots\lambda_{N};\lambda)}=i\frac{\det(\lambda_{j}^{i-1}r_{ij}^{+})_{N}}{\det(\lambda_{j}^{i-1}r_{ij}^{-})_{N}}\psi_{a}(\lambda_{1},\ldots\lambda_{N};\lambda)\,.\label{eq:N-phi_conjugation}
\end{equation}
Plugging the last two formulas in the complex conjugate of (\ref{eq:N-Yaa}),
it follows immediately that $\bar{Y}_{a\dot{a}}=Y_{a\dot{a}}.$ In fact (\ref{eq:N-psi_conjugation})-(\ref{eq:N-phi_conjugation}) imply that the two terms being summed in (\ref{eq:N-Yaa}) are conjugate to each other, so that we can also write
\begin{equation}
Y_{a\dot{a}}=\sqrt{2}\,\text{Re}\left[(1+i)\psi_{a}(\lambda_{1},\ldots\lambda_{N};1)\phi_{\dot{a}}(\lambda_{1},\ldots\lambda_{N};i)\right]\,.
\end{equation}
Moving to the case where $N$ is even, the only extra complication
is that the $r_{ij}^{\pm}$ are also generally phases, $\bar{r}_{ij}^{\pm}=1/r_{ij}^{\pm}$,
see (\ref{eq:N-phi/psi}). The analogues of (\ref{eq:determinant_conjugation})-(\ref{eq:determinat_conjugation2})
will now be,

\begin{equation}\label{eq:determinant_conjugation2}
\begin{aligned}\overline{\det(\lambda_{j}^{i-1}r_{ij}^{\pm})}_{N} & =(-1)^{\frac{N}{2}}\prod_{j=1}^{N}\frac{\psi^{j}}{\phi^{j}}\frac{1}{\lambda_{j}^{N-1}}\det(\lambda_{j}^{i-1}r_{ij}^{\pm})_{N}\,,\\
\overline{\det(\lambda_{j}^{i-1}r_{ij}^{\pm})}_{N+1} & =(-1)^{\frac{N}{2}}\prod_{j=1}^{N+1}\frac{\psi^{j}}{\phi^{j}}\frac{1}{\lambda_{j}^{N}}\det(\lambda_{j}^{i-1}r_{ij}^{\mp})_{N+1},
\end{aligned}
\end{equation}
where we additionally had to take out a factor $\psi^{j}/\phi^{j}$
from the $j$-th column of the conjugated determinant. However the
different factors combine in such a way that the expressions (\ref{eq:N-psi_conjugation})-(\ref{eq:N-phi_conjugation})
remain unchanged, guaranteeing the reality of both $Y_{a\dot{a}}$
and $\alpha(\lambda_{1},\ldots\lambda_{N})$ as before.

Finally, let us note that
\begin{equation}
\frac{W(\phi^{1},\ldots,\phi^{N})}{W(\psi^{1},\ldots,\psi^{N})}=\frac{\det(\lambda_{j}^{i-1}r_{ij}^{+})_{N}}{\det(\lambda_{j}^{i-1}r_{ij}^{-})_{N}}\,,
\end{equation}
which is real due to (\ref{eq:determinant_conjugation}),(\ref{eq:determinant_conjugation2}), and guarantees that the $N$-th dressed sinh-Gordon solution $\alpha(\lambda_{1},\ldots\lambda_{N})$ will also be real if we start with with the $\alpha=0$ vacuum.

In other words the considerations of this appendix justify our choice (\ref{eq:N-phi/psi}) for alternately using single sinh- and sine-Gordon kinks for building the $N$-th dressed solution when $N$ is odd and even respectively. They also offer a verification of the sequences of Crum transformations (\ref{eq:Darboux_sequences}), in particular for $\alpha_N$ with $N$ even in the first line and $N$ odd in the second line. Had we exchanged our choice for the single sinh- and sine-Gordon kinks between $N$ odd and even, we would obtain the $i \beta_N$ solutions of the two sequences, which however don't make sense in this setting from the string point of view, because the quantity $\partial \vec Y\cdot \bar\partial \vec Y$ defining the Pohlmeyer-reduced field will always be real. Presumably these solutions can be mapped to Euclidean worldsheet strings in $R\times S^2$ target space.

%

\nocite{*}
\bibliographystyle{jhep}
\bibliography{multispike}

\providecommand{\href}[2]{#2}\begingroup\raggedright\begin{thebibliography}{10}

\bibitem{Maldacena:2011ut}
J.~Maldacena, {\it {The Gauge/gravity duality}},
  \href{http://xxx.lanl.gov/abs/1106.6073}{{\tt arXiv:1106.6073}}.

\bibitem{Gubser:2002tv}
S.~Gubser, I.~Klebanov, and A.~M. Polyakov, {\it {A Semiclassical limit of the
  gauge / string correspondence}},  {\em Nucl.Phys.} {\bf B636} (2002) 99--114,
  [\href{http://xxx.lanl.gov/abs/hep-th/0204051}{{\tt hep-th/0204051}}].

\bibitem{Gross:1974cs}
D.~Gross and F.~Wilczek, {\it {ASYMPTOTICALLY FREE GAUGE THEORIES. 2.}},  {\em
  Phys.Rev.} {\bf D9} (1974) 980--993.

\bibitem{Georgi:1951sr}
H.~Georgi and H.~Politzer, {\it {Electroproduction scaling in an asymptotically
  free theory of strong interactions}},  {\em Phys.Rev.} {\bf D9} (1974)
  416--420.

\bibitem{Basso:2010in}
B.~Basso, {\it {Exciting the GKP string at any coupling}},  {\em Nucl.Phys.}
  {\bf B857} (2012) 254--334.

\bibitem{Rey:1998ik}
S.-J. Rey and J.-T. Yee, {\it {Macroscopic strings as heavy quarks in large N
  gauge theory and anti-de Sitter supergravity}},  {\em Eur.Phys.J.} {\bf C22}
  (2001) 379--394, [\href{http://xxx.lanl.gov/abs/hep-th/9803001}{{\tt
  hep-th/9803001}}].

\bibitem{Maldacena:1998im}
J.~M. Maldacena, {\it {Wilson loops in large N field theories}},  {\em
  Phys.Rev.Lett.} {\bf 80} (1998) 4859--4862,
  [\href{http://xxx.lanl.gov/abs/hep-th/9803002}{{\tt hep-th/9803002}}].

\bibitem{Alday:2007hr}
L.~F. Alday and J.~M. Maldacena, {\it {Gluon scattering amplitudes at strong
  coupling}},  {\em JHEP} {\bf 0706} (2007) 064,
  [\href{http://xxx.lanl.gov/abs/0705.0303}{{\tt arXiv:0705.0303}}].

\bibitem{Anastasiou:2003kj}
C.~Anastasiou, Z.~Bern, L.~J. Dixon, and D.~Kosower, {\it {Planar amplitudes in
  maximally supersymmetric Yang-Mills theory}},  {\em Phys.Rev.Lett.} {\bf 91}
  (2003) 251602, [\href{http://xxx.lanl.gov/abs/hep-th/0309040}{{\tt
  hep-th/0309040}}].

\bibitem{Bern:2005iz}
Z.~Bern, L.~J. Dixon, and V.~A. Smirnov, {\it {Iteration of planar amplitudes
  in maximally supersymmetric Yang-Mills theory at three loops and beyond}},
  {\em Phys.Rev.} {\bf D72} (2005) 085001,
  [\href{http://xxx.lanl.gov/abs/hep-th/0505205}{{\tt hep-th/0505205}}].

\bibitem{Alday:2007he}
L.~F. Alday and J.~Maldacena, {\it {Comments on gluon scattering amplitudes via
  AdS/CFT}},  {\em JHEP} {\bf 0711} (2007) 068,
  [\href{http://xxx.lanl.gov/abs/0710.1060}{{\tt arXiv:0710.1060}}].

\bibitem{Alday:2009yn}
L.~F. Alday and J.~Maldacena, {\it {Null polygonal Wilson loops and minimal
  surfaces in Anti-de-Sitter space}},  {\em JHEP} {\bf 0911} (2009) 082,
  [\href{http://xxx.lanl.gov/abs/0904.0663}{{\tt arXiv:0904.0663}}].

\bibitem{Alday:2009dv}
L.~F. Alday, D.~Gaiotto, and J.~Maldacena, {\it {Thermodynamic Bubble Ansatz}},
   {\em JHEP} {\bf 1109} (2011) 032,
  [\href{http://xxx.lanl.gov/abs/0911.4708}{{\tt arXiv:0911.4708}}].

\bibitem{Alday:2010vh}
L.~F. Alday, J.~Maldacena, A.~Sever, and P.~Vieira, {\it {Y-system for
  Scattering Amplitudes}},  {\em J.Phys.A} {\bf A43} (2010) 485401,
  [\href{http://xxx.lanl.gov/abs/1002.2459}{{\tt arXiv:1002.2459}}].

\bibitem{Pohlmeyer:1975nb}
K.~Pohlmeyer, {\it {Integrable Hamiltonian Systems and Interactions Through
  Quadratic Constraints}},  {\em Commun.Math.Phys.} {\bf 46} (1976) 207--221.

\bibitem{Scott:1973eg}
A.~C. Scott, f.~Y.~F. Chu, and D.~W. McLaughlin, {\it {The Soliton: A New
  Concept in Applied Science}},  {\em IEEE Proc.} {\bf 61} (1973) 1443--1483.

\bibitem{Barbashov:1980kz}
B.~Barbashov and V.~Nesterenko, {\it {RELATIVISTIC STRING MODEL IN A SPACE-TIME
  OF A CONSTANT CURVATURE}},  {\em Commun.Math.Phys.} {\bf 78} (1981) 499.

\bibitem{DeVega:1992xc}
H.~De~Vega and N.~G. Sanchez, {\it {Exact integrability of strings in
  D-Dimensional De Sitter space-time}},  {\em Phys.Rev.} {\bf D47} (1993)
  3394--3405.

\bibitem{deVega:1992yz}
H.~de~Vega, A.~Mikhailov, and N.~G. Sanchez, {\it {Exact string solutions in
  (2+1)-dimensional de Sitter space-time}},  {\em Theor.Math.Phys.} {\bf 94}
  (1993) 166--172, [\href{http://xxx.lanl.gov/abs/hep-th/9209047}{{\tt
  hep-th/9209047}}].

\bibitem{Larsen:1996gn}
A.~Larsen and N.~G. Sanchez, {\it {Sinh-Gordon, cosh-Gordon and Liouville
  equations for strings and multistrings in constant curvature space-times}},
  {\em Phys.Rev.} {\bf D54} (1996) 2801--2807,
  [\href{http://xxx.lanl.gov/abs/hep-th/9603049}{{\tt hep-th/9603049}}].

\bibitem{Jevicki:2007aa}
A.~Jevicki, K.~Jin, C.~Kalousios, and A.~Volovich, {\it {Generating AdS String
  Solutions}},  {\em JHEP} {\bf 0803} (2008) 032,
  [\href{http://xxx.lanl.gov/abs/0712.1193}{{\tt arXiv:0712.1193}}].

\bibitem{Jevicki:2008mm}
A.~Jevicki and K.~Jin, {\it {Solitons and AdS String Solutions}},  {\em
  Int.J.Mod.Phys.} {\bf A23} (2008) 2289--2298,
  [\href{http://xxx.lanl.gov/abs/0804.0412}{{\tt arXiv:0804.0412}}].

\bibitem{Jevicki:2009uz}
A.~Jevicki and K.~Jin, {\it {Moduli Dynamics of AdS(3) Strings}},  {\em JHEP}
  {\bf 0906} (2009) 064, [\href{http://xxx.lanl.gov/abs/0903.3389}{{\tt
  arXiv:0903.3389}}].

\bibitem{Jevicki:2010yt}
A.~Jevicki and K.~Jin, {\it {AdS String: Classical Solutions and Moduli
  Dynamics}},  \href{http://xxx.lanl.gov/abs/1001.5301}{{\tt arXiv:1001.5301}}.

\bibitem{Janik:2011bd}
R.~A. Janik and A.~Wereszczynski, {\it {Correlation functions of three heavy
  operators: The AdS contribution}},  {\em JHEP} {\bf 1112} (2011) 095,
  [\href{http://xxx.lanl.gov/abs/1109.6262}{{\tt arXiv:1109.6262}}].

\bibitem{Kazama:2011cp}
Y.~Kazama and S.~Komatsu, {\it {On holographic three point functions for GKP
  strings from integrability}},  {\em JHEP} {\bf 1201} (2012) 110,
  [\href{http://xxx.lanl.gov/abs/1110.3949}{{\tt arXiv:1110.3949}}].

\bibitem{Beisert:2010jr}
N.~Beisert, C.~Ahn, L.~F. Alday, Z.~Bajnok, J.~M. Drummond, et~al., {\it
  {Review of AdS/CFT Integrability: An Overview}},  {\em Lett.Math.Phys.} {\bf
  99} (2012) 3--32, [\href{http://xxx.lanl.gov/abs/1012.3982}{{\tt
  arXiv:1012.3982}}].

\bibitem{Drukker:2012de}
N.~Drukker, {\it {Integrable Wilson loops}},
  \href{http://xxx.lanl.gov/abs/1203.1617}{{\tt arXiv:1203.1617}}.

\bibitem{Correa:2012hh}
D.~Correa, J.~Maldacena, and A.~Sever, {\it {The quark anti-quark potential and
  the cusp anomalous dimension from a TBA equation}},
  \href{http://xxx.lanl.gov/abs/1203.1913}{{\tt arXiv:1203.1913}}.

\bibitem{Miramontes:2008wt}
J.~Miramontes, {\it {Pohlmeyer reduction revisited}},  {\em JHEP} {\bf 0810}
  (2008) 087, [\href{http://xxx.lanl.gov/abs/0808.3365}{{\tt
  arXiv:0808.3365}}].

\bibitem{Dorn:2009kq}
H.~Dorn, G.~Jorjadze, and S.~Wuttke, {\it {On spacelike and timelike minimal
  surfaces in AdS(n)}},  {\em JHEP} {\bf 0905} (2009) 048,
  [\href{http://xxx.lanl.gov/abs/0903.0977}{{\tt arXiv:0903.0977}}].

\bibitem{Sakai:2009ut}
K.~Sakai and Y.~Satoh, {\it {A Note on string solutions in AdS(3)}},  {\em
  JHEP} {\bf 0910} (2009) 001, [\href{http://xxx.lanl.gov/abs/0907.5259}{{\tt
  arXiv:0907.5259}}].

\bibitem{Dorn:2009gq}
H.~Dorn, {\it {Some comments on spacelike minimal surfaces with null polygonal
  boundaries in AdS(m)}},  {\em JHEP} {\bf 1002} (2010) 013,
  [\href{http://xxx.lanl.gov/abs/0910.0934}{{\tt arXiv:0910.0934}}].

\bibitem{Ryang:2009ay}
S.~Ryang, {\it {Asymptotic AdS String Solutions for Null Polygonal Wilson Loops
  in R**1,2}},  {\em Mod.Phys.Lett.} {\bf A25} (2010) 2555--2569,
  [\href{http://xxx.lanl.gov/abs/0910.4796}{{\tt arXiv:0910.4796}}].

\bibitem{Sakai:2010eh}
K.~Sakai and Y.~Satoh, {\it {Constant mean curvature surfaces in $AdS_3$}},
  {\em JHEP} {\bf 1003} (2010) 077,
  [\href{http://xxx.lanl.gov/abs/1001.1553}{{\tt arXiv:1001.1553}}].

\bibitem{Dorey:2010iy}
N.~Dorey and M.~Losi, {\it {Giant Holes}},  {\em J.Phys.A} {\bf A43} (2010)
  285402, [\href{http://xxx.lanl.gov/abs/1001.4750}{{\tt arXiv:1001.4750}}].

\bibitem{Ishizeki:2011bf}
R.~Ishizeki, M.~Kruczenski, and S.~Ziama, {\it {Notes on Euclidean Wilson loops
  and Riemann Theta functions}},  \href{http://xxx.lanl.gov/abs/1104.3567}{{\tt
  arXiv:1104.3567}}.

\bibitem{Dorey:2011gr}
N.~Dorey and P.~Zhao, {\it {Scattering of Giant Holes}},  {\em JHEP} {\bf 1108}
  (2011) 134, [\href{http://xxx.lanl.gov/abs/1105.4596}{{\tt
  arXiv:1105.4596}}].

\bibitem{Matveev1991}
V.~Matveev and M.~Salle, {\em {Darboux Transformations and Solitons}}.
\newblock Springer-Verlag, 1991.

\bibitem{Andreev199558}
V.~Andreev and Y.~Brezhnev, {\it Darboux transformation, positons and general
  superposition formula for the sine-gordon equation},  {\em Phys.Lett.} {\bf
  B207} (1995) 58--66.

\bibitem{springerlink:10.1023/A:1007583627973}
H.~Hesheng, {\it Darboux transformations between ${\Delta} \alpha=\sinh \alpha$
  and ${\Delta} \alpha=\sin \alpha$ and the application to pseudo-spherical
  congruences in ${R}^{2,1}$},  {\em Lett.Math.Phys.} {\bf 48} (1999) 187--195.

\bibitem{1999physics...8019C}
M.~M. {Crum}, {\it {Associated Sturm-Liouville systems}},  {\em ArXiv Physics
  e-prints} (1955) [\href{http://xxx.lanl.gov/abs/physics/9}{{\tt physics/9}}].

\bibitem{Zakharov:1973pp}
V.~Zakharov and A.~Mikhailov, {\it {Relativistically Invariant Two-Dimensional
  Models in Field Theory Integrable by the Inverse Problem Technique. (In
  Russian)}},  {\em Sov.Phys.JETP} {\bf 47} (1978) 1017--1027.

\bibitem{Spradlin:2006wk}
M.~Spradlin and A.~Volovich, {\it {Dressing the Giant Magnon}},  {\em JHEP}
  {\bf 0610} (2006) 012, [\href{http://xxx.lanl.gov/abs/hep-th/0607009}{{\tt
  hep-th/0607009}}].

\bibitem{Kalousios:2006xy}
C.~Kalousios, M.~Spradlin, and A.~Volovich, {\it {Dressing the giant magnon
  II}},  {\em JHEP} {\bf 0703} (2007) 020,
  [\href{http://xxx.lanl.gov/abs/hep-th/0611033}{{\tt hep-th/0611033}}].

\bibitem{Jevicki:2007pk}
A.~Jevicki, C.~Kalousios, M.~Spradlin, and A.~Volovich, {\it {Dressing the
  Giant Gluon}},  {\em JHEP} {\bf 0712} (2007) 047,
  [\href{http://xxx.lanl.gov/abs/0708.0818}{{\tt arXiv:0708.0818}}].

\bibitem{Neveu:1977cr}
A.~Neveu and N.~Papanicolaou, {\it {Integrability of the Classical Scalar and
  Symmetric Scalar-Pseudoscalar Contact Fermi Interactions in Two-Dimensions}},
   {\em Commun.Math.Phys.} {\bf 58} (1978) 31.

\bibitem{Giddings:1998yd}
S.~B. Giddings, F.~Hacquebord, and H.~L. Verlinde, {\it {High-energy scattering
  and D pair creation in matrix string theory}},  {\em Nucl.Phys.} {\bf B537}
  (1999) 260--296, [\href{http://xxx.lanl.gov/abs/hep-th/9804121}{{\tt
  hep-th/9804121}}].

\bibitem{Bonora:2002ay}
L.~Bonora, C.~Constantinidis, L.~Ferreira, and E.~Leite, {\it {Construction of
  exact Riemannian instanton solutions}},  {\em J.Phys.A} {\bf A36} (2003)
  7193--7210, [\href{http://xxx.lanl.gov/abs/hep-th/0208175}{{\tt
  hep-th/0208175}}].

\bibitem{Kruczenski:2002fb}
M.~Kruczenski, {\it {A Note on twist two operators in N=4 SYM and Wilson loops
  in Minkowski signature}},  {\em JHEP} {\bf 0212} (2002) 024,
  [\href{http://xxx.lanl.gov/abs/hep-th/0210115}{{\tt hep-th/0210115}}].

\bibitem{WADATIMiki:1975-02-25}
M.~Wadati, H.~Sanuki, and K.~Konno, {\it Relationships among inverse method,
  backlund transformation and an infinite number of conservation laws},  {\em
  Prog.Theor.Phys.} {\bf 53} (1975) 419--436.

\bibitem{Sommerfield:2008hu}
C.~M. Sommerfield and C.~B. Thorn, {\it {Classical Worldsheets for String
  Scattering on Flat and AdS Spacetime}},  {\em Phys.Rev.} {\bf D78} (2008)
  046005, [\href{http://xxx.lanl.gov/abs/0805.0388}{{\tt arXiv:0805.0388}}].

\bibitem{Berkovits:2008ic}
N.~Berkovits and J.~Maldacena, {\it {Fermionic T-Duality, Dual Superconformal
  Symmetry, and the Amplitude/Wilson Loop Connection}},  {\em JHEP} {\bf 0809}
  (2008) 062, [\href{http://xxx.lanl.gov/abs/0807.3196}{{\tt
  arXiv:0807.3196}}].

\bibitem{1977PhRvB..15.3353L}
G.~{Leibbrandt}, {\it {Exact solutions of the elliptic sine equation in two
  space dimensions with application to the Josephson effect}},  {\em Phys.Rev.}
  (1977) 3353--3361.

\bibitem{Kalousios:2010ne}
C.~Kalousios and G.~Papathanasiou, {\it {Giant Magnons in Symmetric Spaces:
  Explicit N-soliton solutions for $CP^n, SU(n)$ and $S^n$}},  {\em JHEP} {\bf
  1007} (2010) 068, [\href{http://xxx.lanl.gov/abs/1005.1066}{{\tt
  arXiv:1005.1066}}].

\bibitem{Rajaraman}
R.~Rajaraman, {\em {Solitons and Instantons:An Introduction to Solitons and
  Instantons in Quantum Field Theory}}.
\newblock North-Holland, 1987.

\end{thebibliography}\endgroup

\end{document}